\documentclass[A4,12pt]{article} 
\usepackage{multirow} 
\usepackage{graphicx}
\usepackage{xcolor}             
\usepackage{amssymb}
\usepackage[round,authoryear]{natbib}
\usepackage{hyperref}
\hypersetup{colorlinks=true,linkcolor=blue,citecolor=blue,breaklinks={true}}

\font\bbfnt=msbm10

\def\bbR{\mbox{\bbfnt R}}
\def\bbN{\mbox{\bbfnt N}}

\newcommand{\T}{\scriptscriptstyle\rm T }
\newcommand{\bxi}{\mbox{\boldmath $\xi$}}
\newcommand{\bpsi}{\mbox{\boldmath $\psi$}}
\newcommand{\btheta}{\mbox{\boldmath $\theta$}}

\newcommand{\bu}{\mbox{\boldmath $u$}}
\newcommand{\by}{\mbox{\boldmath $y$}}
\newcommand{\be}{\mbox{\boldmath $e$}}

\newtheorem{lemma}{Lemma}
\newtheorem{theorem}{Theorem}
\newtheorem{definition}{Definition}

\newcommand{\m}[0]{\textcolor{blue}{${\cal M}$}~}
\newcommand{\s}[0]{\textcolor{red}{${\cal S}$}~}
\definecolor{green}{rgb}{0,0.6,0}
\newcommand{\I}[0]{\textcolor{green}{${\cal I}$}~}

\begin{document}
\begin{center}

{\bf  A Bird's Eye View of Nonlinear System Identification}\vspace{0.5cm}

Luis A. Aguirre \vspace{0.5cm}

{Departamento de Engenharia Eletr\^onica \\
Programa de P\'os-Gradua\c{c}\~ao em Engenharia El\'etrica\\
Universidade Federeal de Minas Gerais ---  
Av. Ant\^onio Carlos 6627, 31270-901 Belo Horizonte MG, Brazil\\
{\tt aguirre@ufmg.br}}

\end{center}

\begin{abstract}
This text aims at providing a bird's eye view of system identification with special
attention to nonlinear systems. The driving force is to give a feeling for the philosophical 
problems facing those that build mathematical models from data. Special attention will
be given to grey-box approaches in nonlinear system identification. In this text, 
grey-box methods use auxiliary information such as the system steady-state data,  
possible symmetries, some bifurcations and the presence of hysteresis.
The text ends with a sample of applications.
No attempt is made to be thorough nor to survey such an extensive and
mature field as system identification. In most parts references will be provided for a 
more detailed study.
\end{abstract}

\tableofcontents

\section{Introduction}
\label{intro}

System identification is the ``art'' of building dynamical mathematical models from data, which
are measured from a system which, in principle, could be any dynamical system. 
A typical system identification problem can be divided into five steps: i)~testing and
data collection, ii)~choice of model class, iii)~structure selection, iv)~parameter estimation
and v)~model validation. One of the aims of this text is to provide some preliminary
discussion to each of these steps, with a clear bias towards nonlinear system identification.
Black-box and  grey-box techniques will be mentioned.

In doing this, a somewhat ``philosophical framework'' will be proposed in order to 
bring home to the newcomer some of the real challenges of this fascinating field. Such a
framework is admittedly subjective, but it is believed that it will prove helpful in understanding
a few important problems and fundamental challenges in system identification.

\begin{figure}[!ht]
		\centering
		\includegraphics[width=0.80\textwidth]{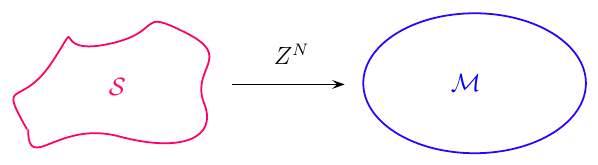}
		\caption{\small Simplified schematic diagram for black-box identification, where \s represents the system
		that should be approximated by a model \m which is built from a set of
		measured data $Z^N$ of length $N$.}
		\label{f310119a}
\end{figure}

The statement of a system identification problem is simple and can be declared thus (see Figure~\ref{f310119a}): given a dynamical system \s \hspace{-0.15cm},
possibly nonlinear, from which a set of measured data $Z^N$ is available, find a mathematical model \m 
that represents \s in some meaningful way. To build \m exclusively from $Z^N$ is a {\it black-box identification problem}. 
In {\it grey-box problems}, besides $Z^N$ there will be some other source of information about \s \hspace{-0.15cm}.
In $Z^N$, $N$ indicates the length of the data set and will be omitted in general.\footnote{Some authors restrict
grey-box problems to that of estimating parameters of an equation that comes from first principles.}

The following points should be noticed:

\begin{enumerate}
\item in black-box system identification, because $Z$ is all there is to build \m \hspace{-0.15cm},
all relevant information about \s should be in such data. Hence, $Z$ has to be carefully obtained;

\item the model \m must be a member of a model class ${\cal M}^{\rm c}$ that should be
consistent with the relevant aspects of \s that are aimed at. For instance, if \s is
dynamical and nonlinear, so must be ${\cal M}^{\rm c}$;

\item given  a class of models ${\cal M}^{\rm c}$ consistent with \s and the modelling aims, there are many specific candidate
models, i.e. $\{ {\cal M}_1,\,{\cal M}_2 \ldots {\cal M}_n\}\in {\cal M}^{\rm c}$ which are not  equivalent
nor necessarily adequate to represent specific features of \s \hspace{-0.15cm};

\item a parametric model \m is composed by variables and parameters which must be chosen and estimated;

\item a mathematical model \m is not directly comparable to a physical system \s \hspace{-0.15cm}, because 
they are entities of different natures. How can one decide if \m represents \s \hspace{-0.13cm}?

\end{enumerate}

The five aspects just mentioned are closely related to the five steps of a typical identification problem, 
listed in the opening paragraph. In general terms, this text will be organised in two parts.
First, each of the aforementioned five steps will be briefly introduced. This is done in Sections~\ref{s1} to~\ref{s5}.
In the second part, some specific aspects, especially related to grey-box techniques will be presented
in Section~\ref{gb}. Section~\ref{eps} discusses three case studies in order to illustrate some of the main
ideas. The work concludes with suggestions for further reading.

\section{Testing and Data Collection}
\label{s1}

There are a number of situations in system identification in which one must build the model
from historical data. By historical data it is meant data that have been collected not as a
result of any specific designed test. For now, we will assume that a specific test can be 
performed to produce $Z$. Before we start it will be convenient to point out that the data
are very often composed by inputs $\bu$ and outputs $\by$. For now, we will restrict ourselves
to the SISO (single-input single-output) case, hence $Z^N=[u(k)~y(k)],~k=1,\,2, \ldots N$.

The standard situation in system identification is that both \s and therefore \m are dynamical.
Hence, it is necessary that $Z$ be produced as a result of a dynamical test on \s \hspace{-0.15cm}.
That is, the input $u(k)$ is designed in such a way that the relevant dynamics of \s appear in $y(k)$.
A fundamental point here is to realize that because \m is built exclusively from $Z$ (the black-box case),
features of \s that do not show up in $Z$ will most likely not appear in \m either. In discussing this
point, it will be convenient to address the linear and nonlinear cases in turn.

\subsection{Testing}

Every physical system \s is typically nonlinear and time-varying. Let us assume that the intended
model \m is linear. This means to say that we are interested in the linear aspects of the dynamical 
behaviour of \s \hspace{-0.15cm}. Hence $Z$ must be consistent with such an aim. In order to 
guarantee this and avoid that nonlinear features of \s appear in $Z$, a typical test is to excite 
\s around an operating point, that is to say, that the amplitude range of $u(k)$ must be limited
to a region in which ``the nonlinearities of \s are not excited''. 

Also, because \m is intended to be 
dynamical, $Z$ must have such information. If $u(k)$ is too slow, the system \s will not have any
troubles in following the input and the dynamics will not appear in $y(k)$. When a driver slows
down before passing a speed bump the idea is {\it not}\, to excite the dynamics of the vehicle
suspension. So, in general, slow input signals result in data with poor dynamical information. 
On the other extreme the practitioner also faces problems. If the input is too fast, then there is
not enough time for the system to react to the input changes and there is no significant energy
transfer from the input to the system. In such a situation the data also turn out to be poor
from a dynamical point of view. In technical terms, the power spectrum of the input
$\Phi_u(\omega)={\cal F}(R_u)$\footnote{${\cal F}$ denotes de Fourier transform and $R_u$
is the covariance matrix of $u(k)$. The power spectrum of a signal is defined as the Fourier transform of the covariance
matrix of such a signal. } must have sufficient power in the frequency range of
interest, that is, in the frequency range of the dominant dynamics of \s \hspace{-0.15cm}.
If $\Phi_u(\omega)$ is nonzero at $n$ frequencies, then $u(k)$ is
said to be {\it persistently exciting of order n}. In practice, a signal is said to be
persistently exciting if it is sufficiently rich in order to facilitate the estimation process.

Because \m should be time-invariant and \s is time-varying, one should guarantee that
the time-varying aspect does {\it not}\, appear in $Z$. This is typically achieved by performing
dynamical tests that are not too long. In other words, the changes in \s during the test should be
negligible. The newcomer to the theory of dynamical systems should be aware that supposing
a time-invariant system does {\it not}\, imply a constant output.

What changes if we aim at a nonlinear model? In what concerns the time-varying aspect of \s
nothing changes because we still aim at a time-invariant model \m \hspace{-0.15cm}. Hence the
period during which the data $Z$ are collected still needs to be sufficiently short as to guarantee
that it is reasonable to consider that \s did not change significantly during that period.

Because \m is now nonlinear, then it is required that the relevant nonlinear aspects of \s be
present in $Z$. Clearly, the test cannot possibly be performed as for the linear case, on the
contrary, a typical test for nonlinear system identification will probably specify large variations
in the input signal $u(k)$ in order to excite the nonlinearities of \s and guarantee that they
appear in the data $Z$. From a practical point of view, it is often difficult and unsafe to drive
the system \s over a wide range of operating conditions. This is one of the practical challenges
faced by the practitioner that aims at building nonlinear models from data. Fortunately, there
are possibilities that help to face such challenges. A common one is to perform several low amplitude
tests over a set of operating points that cover the region of interest. The inconvenience of this
is that the test can turn out to be long. In Section~\ref{subssd} we will discuss another solution
to this problem based on grey-box techniques. 

If from an amplitude point of view the testing of nonlinear systems is more challenging than
for linear systems, in terms of frequency content of the input, nonlinear systems are somewhat 
easier. The reason for this is that nonlinear systems transfer spectral power among different
frequencies, which does {\it not}\, happen in linear systems. To see this, suppose we choose
as input $u(k)=A\cos (\omega_0 t)$ with a sufficiently small value of $A$ in order not to excite
the nonlinearities in \s \hspace{-0.15cm}. Linear systems theory tells us that in steady-state
the output  will be of the form $y(k)=a_0 \cos (\omega_0 t + \phi_0)$. In other words, the
gain {\it at frequency $\omega_0$}\, is $a_0/A$ and the phase at that frequency is $\phi_0$
(usually a negative value). That is we have identified the Bode diagram (frequency response)
of \s only at $\omega_0$. Now it should become clear why $u(k)$ should be persistently exciting
of large order $n$: to have information at $n$ different frequencies.

\begin{figure}
\centering
\includegraphics[scale=0.5]{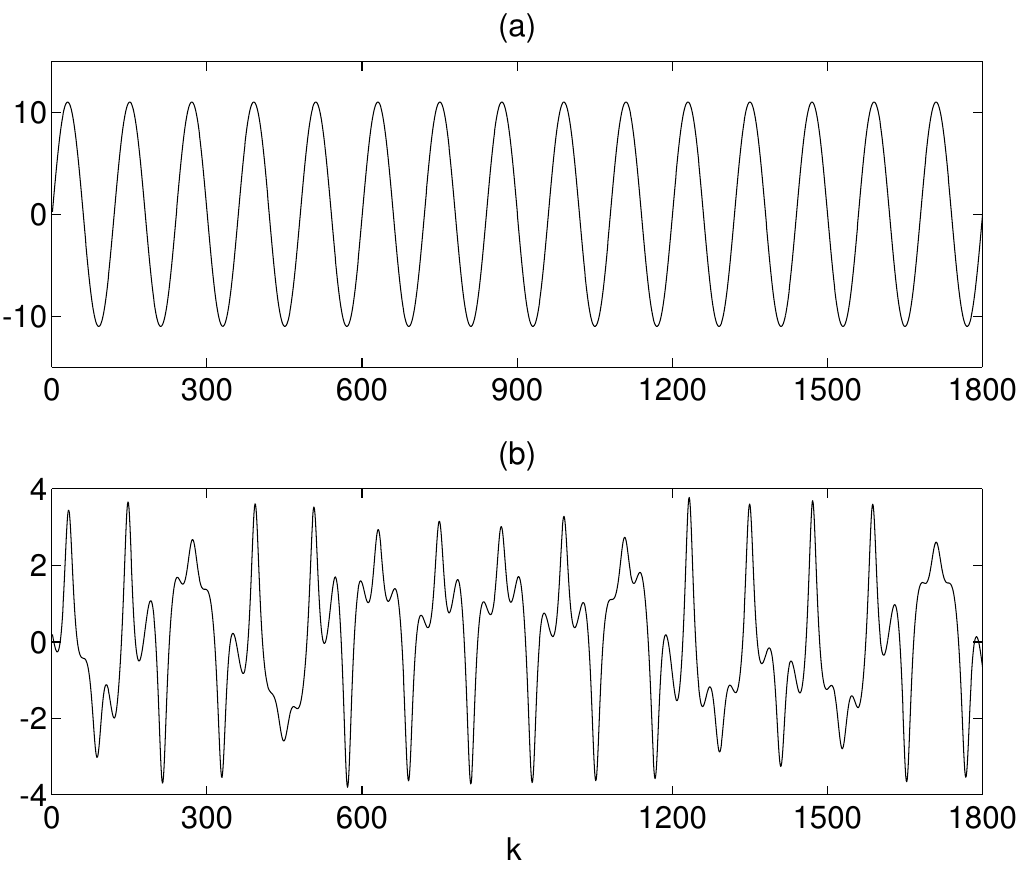} 
\caption{\small \label{duffued1} Simulated data from the Duffing-Ueda oscillator
$\ddot{y} + 0.1\dot{y} + y^3 = u(t) $ for input (a)~$u(t)=11\cos (t)$, and (b)~output $y(t)$.
Notice how $y(k)$ has many other frequencies besides $\omega=1$\,rad/s. That is
due to nonlinearity.}
\end{figure}

Suppose that $A$ is increased in order to excite the nonlinearities. Figure~\ref{duffued1} shows
both the input and output for this case. Clearly, the output has more than one frequency. This is
a direct consequence of nonlinearity. Hence, in the case of nonlinear systems there is somewhat
less strain on the test in what concerns frequency content of the input. A simple summary of this
discussion is: for linear models input amplitudes should be low and the spectrum should be wide;
for nonlinear models the amplitude profile should be large whereas the input spectrum can 
sometimes be narrow.

As a final remark in this section, what if we only have historical data in hand? The same principles
apply, but now instead of performing a test to measure $Z$ with the desired features,
one must go through all the available data and find windows that present the desired
characteristics. Such windows are candidates to compose $Z$. Procedures for detecting 
transients from a historical record can be found in \citep{rib_agu/15,bit_eal/15}.

\subsection{Choosing the sampling period}

When it comes to choosing the sampling time $T_{\rm s}$, one immediately thinks in terms of
Shannon's sampling theorem: a signal that does {\it not}\, have any components of frequency 
above $f_{\rm max}=1/2T_{\rm s}$ can be unambiguously reconstructed from a set of samples regularly
spaced in time by $T_{\rm s}$. In many practical problems, including system identification, this
result is not totally practical for a couple of reasons. First and foremost, $f_{\rm max}$ is generally not known
beforehand, second, to sample a signal with a frequency just above  $2f_{\rm max}$ is a lower
bound rather than a comfortable working value.

\begin{sloppypar}
On the other hand someone might suggest oversampling the data. This also has its troubles as
consecutive samples are highly redundant and therefore the numerical problems that must be solved
typically become ill-conditioned. A practical procedure that works well in many situations is the following.
First, using an admittedly short sampling time record an oversampled data set $Z^*=[u^*(k)~y^*(k)],~k=1,\,2,
\ldots$. Now we want to choose a decimation factor $\Delta \in \bbN$ such that $u(k)=u^*(\Delta k)$
and $y(k)=y^*(\Delta k)$. It is assumed that all signals are band-limited. 
\end{sloppypar}

First, the following covariance functions are computed
\begin{eqnarray}
\label{deltacor}
r_{y^*}(\tau) & = & {\rm E}\left[ (y^*(k)-\overline{y^*(k)}) 
		 (y^*(k-\tau )-\overline{y^*(k)}) \right] \nonumber \\
r_{y^{*2'}}(\tau) & = & {\rm E}\left[ (y^{*2}(k)-\overline{y^{*2}(k)}) 
		 (y^{*2}(k-\tau )-\overline{y^{*2}(k)}) \right]  ,
\end{eqnarray}

\noindent
where E[$\cdot$] indicates the mathematical expectation and the overbar indicates time-average.
The first expression in (\ref{deltacor}) is the standard linear covariance function, whereas the second
is a nonlinear function. Notice that such functions are computed using the oversampled data.

Second, plot functions $r_{y^*}(\tau) $ and $r_{y^{*2'}}(\tau)$. Call $\tau_{y^*}$  $\tau_{y^{*2'}}$ the
lags at which the first minimum of each function occurs. Choose the least of them, that is
$\tau_{\rm m}^*={\rm min}[ \tau_{y^*},~\tau_{y^{*2'}}]$. Call $\tau_{\rm m}$ the corresponding value
for the data decimated with factor $\Delta$. Hence, we wish to determine $\Delta$ such that 
\begin{equation}
\label{criteriotau}
 10 \le \tau_{\rm m} \le 20  ,
\end{equation}

\noindent
where the limits can sometimes be relaxed to 5 and 25.

\begin{figure}
\centering
\begin{tabular}{cc}
  (a) & (b) \\  
\includegraphics[scale=0.3]{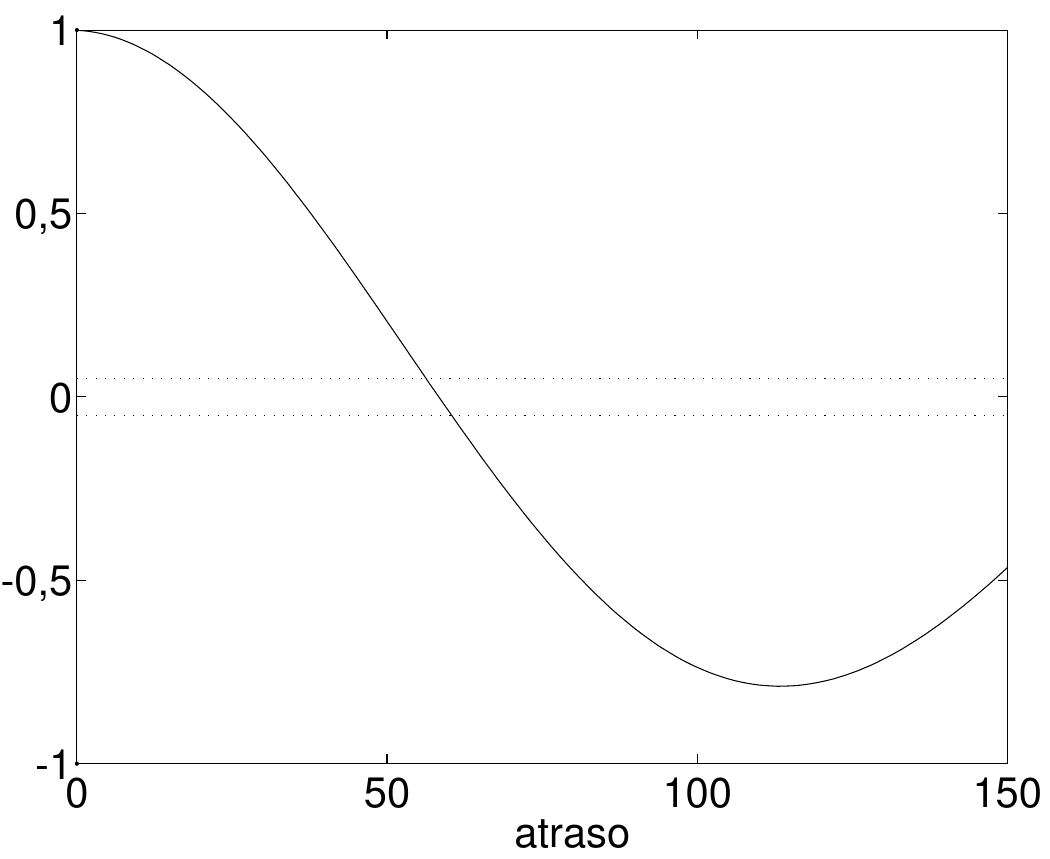} &
\includegraphics[scale=0.3]{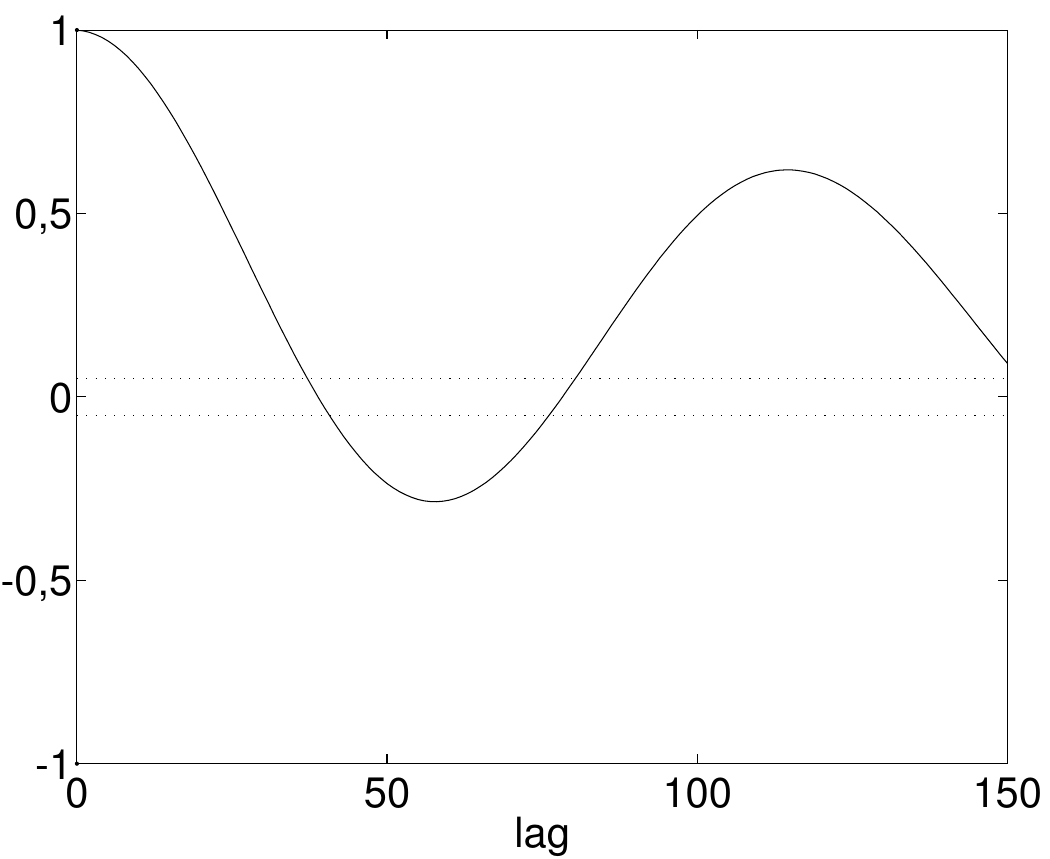} 
\end{tabular}
\caption{\small \label{dscorr}Autocovariance functions for an oversampled signal $y^*(k)$:
(a)~$r_{y^*}(\tau)$ and (b)~$r_{y^{*2'}}(\tau)$. In this example  $\tau_{\rm m}^*={\rm min}[ \sim 115,~\sim 55]=55$.
Plot (a)~was obtained using the $x$ variable of Chua's oscillator, and plot (b)~with the $y$ variable of
the same system. This shows that an acceptable samplint time for the $x$ variable would result in
undersampling the $y$ variable. See \citep{agu/95} for details.}
\end{figure}

For example, consider an oversampled signal $y^*(k)$ for which the linear and nonlinear covariance
functions are shown in Figure~\ref{dscorr}. The smallest lag corresponding to the first minimum is
$\tau_{\rm m}^*=55$, hence if we choose $\Delta = 4$, this value for the decimated signal will be
$\tau_{\rm m} \approx 14$, which satisfies (\ref{criteriotau}). Hence the decimation factor could be
$\Delta = 4$. As a final remark, it should be carefully noticed that if the data $Z$ is composed of more
than one signal, the decimation factor applied must be the same for all signals. Therefore, the $\Delta$ 
used must be the smallest that will satisfy the decimation criterion for all signals.

Some technical aspects of testing can be found in \citep{leo_bil/87,gev_eal/09} and specifically for
closed-loop systems in \citep{bom_eal/06}. The decimation criterion discussed in this section was
originally put forward in \citep{agu/95} and some effects on aspects of system identification have been
discussed in \citep{bil_agu/94}.

\section{Choice of Model Class}
\label{s2}

This is probably the step which receives less attention and it is not really difficult to see why. From
a more technical point of view, it suffices that the model class ${\cal M}^{\rm c}$ be sufficiently general
to include the relevant aspects of \s \hspace{-0.15cm}. From a practical point of view, the practitioner
tends to use the model class that he/she is familiar with. Can we be more specific and ``scientific'' in
choosing the model class?

Probably the most important choice is to decide if ${\cal M}^{\rm c}$ will be linear or not. This of course
has to do with the intended use of the model. For instance, although linear models are admittedly less
effective in representing systems in general, they could be preferred, say, for control purposes. Within
the class of linear models, there are some subclasses as, for instance, transfer functions and state space 
representations. If a model is intended to implement a Kalman filter, then ${\cal M}^{\rm c}$ is likely
to be the class of linear models in state-space form \citep{van_dem/96,bor_gar/11}.

In the realm of nonlinear model classes, the variety is very large, in what follows only a few are mentioned:
Volterra series \citep{cam_eal/06}, nonlinear output frequency response functions \citep{bay_eal/18},
Hammerstein and Wiener models \citep{bai/02,agu_eal/05iee,sch_tie/17}. These model classes have been surveyed
in \citep{bil/80}. Polynomial and rational NARMAX (nonlinear autoregressive moving average models with
exogenous variables) \citep{leo_bil/85a,bil_che/89b,bil_zhu/91}. A recent overview of such 
model classes and related methods can be found in \citep{bil/13book}. Differential equations (continuous-time
models) have been used in \citep{gou_let/94,man_eal/12}, radial-basis functions \citep{che_eal/90rbf,oga_eal/96}
and neural networks \citep{nar_par/90,che_eal/90nn,bil_che/92,bal_gom/02,for_pig/21}. See \citep{roc_ser/17} 
and references therein for a description of NARX neuro-fuzzy models. The main motivation for 
using such nonlinear representations is that they are universal approximations, hence regardless of
the features of \s \hspace{-0.15cm}, such model classes are sufficiently general to represent the system.

The model classes mentioned in the previous paragraph can be classified as global in the sense that a
single model structure is used to map the whole set of inputs to the outputs. On the other extreme, one
finds models formed by a combination of linear models that are responsible for representing the
data only locally \citep{bar_sou/16,bar_eal/18,liu_eal/19}. See \citep{joh_fos/93,got_eal/98} for some
early works on this topic. An intermediary class is that of regional models \citep{jun_eal/15}.

Although several of such representations can be considered fairly general, the difficulty of using them in
practical problems may vary greatly, as will be pointed out later. Also, in Section~\ref{gb} it will be argued
that the model class can be chosen based on the ease with which auxiliary information can be used in 
the process of model building. This will prove to be one of the few objective criteria available for choosing
the model class.

In order to facilitate discussions about structure selection and parameter estimation, it will be convenient
to choose a working model class. Although most of what will be said is applicable to any model class, 
in what follows we shall consider the class of NARMAX models \citep{che_bil/89}
\begin{eqnarray}
\label{NARMAX}
   y(k) & = & F^\ell [ y(k-1),\ldots,y(k-n_y),u(k-d) \! \ldots \nonumber \\
    & &  \! u(k-n_u),e(k-1), \ldots e(k-n_e), e(k) ]  ,
\end{eqnarray}
where  $e(k)$ accounts for uncertainties and possible noise. Also $n_y,~n_u$ and $n_e$ are the maximum lags in each variable, and $d\ge 1$ is the pure delay. The function $F$ can be any nonlinear function, such as a 
neural network, radial basis function, rational or polynomial, each of which determines a
different model class. For the sake of discussion, in what follows
$F^\ell [\cdot]$  will be a polynomial  with nonlinearity degree $\ell$. To estimate the pure
delay $d$ is both important and challenging in the context of system identification \citep{alv_eal/17}.

\section{Structure Selection}
\label{s3}

Having decided which model class to use, it is important to realize that there are scores
of different model structures within the chosen class ${\cal M}^{\rm c}$, that is 
$\{ {\cal M}_1,\,{\cal M}_2 \ldots {\cal M}_n\}\in {\cal M}^{\rm c}$,
where $n$ can be very large. On its own that remark justifies the search for a much smaller
set of model structures that should be considered in a given situation. However the problem is
indeed far more critical than what it seems at first sight. If the model class ${\cal M}^{\rm c}$ is
very general, the temptation to grow the model more than needed is great. This is known in
the literature as {\it overparametrization}\, or {\it overfitting}\, and has severe detrimental effects on the 
dynamics of the identified models \citep{agu_bil/93b}. In this section we will provide a brief
discussion for model classes that are linear-in-the-parameters to give the reader a feel for the problem.
In the next section it will be explained why it is relatively ``natural'' to overparametrize a model.

We start by assuming that ${\cal M}^{\rm c}$ is the class of single-output NARX polynomials. 
Hence, the corresponding model -- see (\ref{NARMAX}) --  can rewritten as:
\begin{eqnarray}
\label{NARX}
   y(k) & = & \theta_1 \psi_1 + \theta_2 \psi_2 + \ldots + \theta_{n_\theta} \psi_{n_\theta} + e(k) \nonumber \\
   & = & [\psi_1~\psi_2 ~\ldots ~\psi_{n_\theta}] \btheta +e(k) = \bpsi (k-1)^{\T} \btheta +e(k) 
\end{eqnarray}

\noindent
where $\psi_i$ is the $i$th regressor and $\theta_i$ the corresponding parameter which
can be written in vector form as $ \btheta=[\theta_1~\theta_2 ~\ldots ~\theta_{n_\theta}]^{\T}$. 
The (column) vector of regressors is $\bpsi (k-1)$, indicating that all regressors
are taken at most up to instant $k-1$. The 
regressors are any combinations up to degree $\ell$ of input and/or output variables
down to lags $n_u$ and $n_y$, respectively. For instance, $\psi_1=y(k-1)$, 
$\psi_2=u(k-3)$ and $\psi_3=y(k-1)u(k-2)^2$ are possible regressors for $n_y=n_u=\ell=3$.

Hence, the regressors of model  (\ref{NARX}) may contain any combination
of lagged inputs, outputs and noise terms up to degree $\ell$. The number of such combinations
is determined by the values of $\ell$, $n_y,~n_u$ and $n_e$
and can easily include thousands of candidate regressors. This huge
amount of terms is a major impediment to the usefulness of the 
estimated model and some kind of mechanism is called for in order to 
automatically choose the best $n_\theta$ regressors to compose the model. 
This problem is often referred to as {\it model structure selection}\, and must be
judiciously accomplished regardless of the mathematical representation being used.

This model class is said to be linear-in-the-parameters because all the known parameters
can be separated from the known regressors, as seen in (\ref{NARX}). In the 
case of model classes that are not linear in the parameters, typically we would find
unknown parameters in $\bpsi(k-1)$. The fact that a model be linear-in-the-parameters does
not mean that it satisfies the superposition principle, on the contrary. Many linear-in-the-parameters
model classes are strongly nonlinear.

\subsection{The ERR, SRR and SSMR criteria}
\label{errsrr}

Because of their usefulness and wide acceptance, two criteria for choosing the regressors
of NARX polynomial models are briefly described next. 

A widely used criterion in the structure selection is the {\it error reduction ratio}\, 
(ERR) \citep{bil_eal/89}. This criterion, which is based on one-step-ahead 
prediction error minimization, evaluates the importance of 
the model terms according to their ability to explain the output variance.

The reduction in the variance of the residuals, that occurs as
new terms are included in the model, can be normalized in relation to the 
output variance $\sigma_y^2$. Then, the error reduction ratio due to the inclusion of the $i$th 
regressor in the model can be written as:
\begin{eqnarray}
\label{eq_errp}
{\rm ERR}_i=\frac{ {\rm MS1PE} ({\cal M}_{i-1}) - {\rm MS1PE} ({\cal M}_{i}) }{\sigma_y^2}, 
\hspace{0.5cm}  i=1,\,2,\, \ldots, \,n,
\end{eqnarray}

\noindent
where ${\rm MS1PE} ({\cal M}_{i})$ stands for the mean square one-step-ahead (OSA) prediction error of
the model with $i$ terms (regressors); $n$ is the number of candidate terms tested for; and
${\cal M}_i$ represents a family of models with nested structures, thus ${\cal M}_{i-1}  \subset {\cal M}_{i}$.
In (\ref{eq_errp}) the numerator equals the reduction in variance of the residuals due
to the inclusion of the $i$th regressor. 

A somewhat related criterion, called {\it simulation error reduction ratio} (SRR), was defined in \citep{pir_spi/03} as:
\begin{eqnarray}
\label{eq_srr}
{\rm SRR}_i=\frac{ {\rm MSSE} ({\cal M}_{i-1}) - {\rm MSSE} ({\cal M}_{i}) }{\sigma_y^2}, 
\hspace{0.5cm} i=1,\, 2, \, \ldots, \,n,
\end{eqnarray}

\noindent
where ${\rm MSSE} ({\cal M}_{i})$ stands for the mean square simulation error of
the model with $i$ terms (regressors). In (\ref{eq_srr}) the {\it free-run}\, simulation is used.
In Sec.\,\ref{s4} the important difference between OSA prediction and simulation errors will be
pointed out.

In the same vein, the {\it simulation similarity maximization rate}\, was proposed as \citep{ara_eal/19}:
\begin{eqnarray}
\label{eq_ssmr}
{\rm SSMR}_i=\frac{ \hat{V}_\sigma (y,\, \hat{y}_{i+1}) - \hat{V}_\sigma (y, \,\hat{y}_i) }{\sigma_y^2}, 
\hspace{0.5cm} i=1,\, 2, \, \ldots, \,n,
\end{eqnarray}

\noindent
where $\hat{y}_i$ is the free-run simulated output of the current model and  $ \hat{y}_{i+1}$ refers to 
the model that has the same regressors as the one that produced $\hat{y}_i$ with the addition of the regressor 
that is being tested and $n$ is the number of candidate regressors. 
In (\ref{eq_ssmr}) $\hat{V}_\sigma (X,\,W)$ is the correntropy between the random
variables $X$ and $W$ and quantifies the average similarity between them.

The SRR, is effective in non ideal identification 
conditions and often yields more compact models. On the other hand, such a criterion 
requires a significantly greater computational effort as it will be discussed in Section~\ref{s4}.
To partially circumvent such a limitation alternative procedures have been proposed \citep{bon_eal/10,far_pir/10}.
The SSMR, as the SRR, benefits from using free-run simulated model outputs. Also, it is less
sensitive to eventual problems caused by non-Gaussianity which is the norm in the nonlinear context.

See \citep{men_bil/01,wei_eal/04,hon_eal/08,pir/08,wei_bil/08,alv_eal/12,mar_eal/13,gu_wei/18} for a comparison of methods and
some recent techniques on structure selection approaches somewhat related to the ERR,
SRR and SSMR methods.

\subsection{Other criteria}

Twin concepts that were developed to aid in structure selection problems are the {\it term clusters},
indicated by $\Omega{y^p u^{m}}$, and the respective {\it cluster coefficients}, by $\Sigma {y^p u^m}$. 
Terms of the form $y^p(k-\tau_j)u^m(k-\tau_i) \in \Omega_{y^p u^m}$ for \hbox{$m \! + \! p \leq \ell$}, 
where $\tau_i$ and $\tau_j$ are any time lags. For instance, for the model
$y(k)=\theta_1y(k-1)y(k-2) + \theta_2y(k-1)u(k-2) + \theta_3y(k-3)u(k-3)$
we have $n_y=n_u=3,\,d=2,\,\ell=2$. This model has
two term clusters, namely: $\Omega_{y^2}$ with coefficient $\Sigma_{y^2}=\theta_1$
and $\Omega_{uy}$ with coefficient $\Sigma_{uy}=\theta_2+\theta_3$. 

Often the coefficients of spurious term clusters become very small or oscillate around zero as the model increased in size, hence this could be used to aid in detecting the order of linear models
\citep{agu/94b} or to detect and discard term clusters in nonlinear system identification \citep{agu_bil/94b,agu_eal/97}.
Later on, these concepts turned out to be very useful in grey-box identification problems, was will be
discussed later. 

The use of correlation test has been discussed in \citep{sto_eal/86,leo_bil/87a}, but the procedure is less
specific than what one would like.

In {\it semiphysical modeling}\, prior knowledge of the system is used to establish suitable, though usually
nonlinear, terms of the measurements in order to improve on the model structure \citep{lin_lju/95}. In other words,
in {\it semiphysical modeling}\, physical insight of the system being modeled is used to determine key term clusters.
Of course, {\it semiphysical modeling}\, falls into the category of grey-box techniques.

Following a probabilistic framework, the Randomized algorithm for Model Structure Selection (RaMSS) was proposed recently \citep{fal_eal/15}. The method was formulated for the NARX polynomial model class and extended to cope with 
NARMAX polynomial model class in \citep{ret_agu/19}. The main idea of this approach is to start with a tentative
probability distribution over regressors, e.g. a uniform distribution indicating that all regressors start having the
same probability of composing the model. Using such distribution sample models are produced and tested. 
Regressors that compose good models have their probability increased whereas regressors of poorly performing
models have a smaller probability of being chosen in future samples after several iterations, the best
regressors have a high probability whereas bad regressors have small or even null probability. In the original
RaMSS algorithm the regressors are dealt with independently. A more refined approach that takes into account
possible relationships among regressors has been developed in \citep{bia_eal/17}. In short, in the last reference
the inclusion of a regressor in a model is based on the previously selected regressors. This does not mean 
that the model is built in a forward regression style, because regressors can be discarded from the model.

The bias/variance dilemma, originally discussed in the context of neural networks \citep{gem_eal/92},
is useful for understanding the pros and cons of {\it slight}\, overparametrization. Such a dilemma underlines
several so called {\it information criteria}\, such as Akaike's criterion \citep{aka/74}. These criteria are
helpful to decide the size of a model. Hence, ERR, SRR and SMRR criteria help rank regressors according to their
importance but usually a different criterion is needed to decide where to stop including terms into the model.
A very interesting feature has been reported concerning the bias/variance dilemma for network models
\citep{bel_eal/19}. In short, strongly overparameterized models -- models with a number of features
(independent terms) that greatly exceeds the number of data in the training set -- eventually experience a
second decrease in the mean squared prediction error. This is called the second ``descent'' and has not
only been verified in some classification problems but also in modeling dynamics \citep{rib_eal/21}. The
generality of such results still needs to be established but so far they seem to apply in network-type models
for which it is viable to easily increase the number of features used by the model.

By explicitly including the number of model terms, thus increasing the learning dimension, Hafiz and co-workers 
have used particle swarm algorithms to address the problem of structure selection in nonlinear system
identification \citep{haf_eal/19a}.

The problem of determining the model structure clearly extends to other model classes. For instance, in the
case of neural networks there is an extensive literature in {\it pruning}\, methods \citep{ree/93}. The ERR
has been adapted for choosing the centres of radial basis function networks \citep{che_eal/90rbf,agu_eal/07}.
As it will be discussed in Sec.\,\ref{gb}, auxiliary information regarding fixed points and steady-state at large
provides important clues  for structure selection in the case of polynomial models.

In closing this section it is important to point out that for a set of data of limited size and quality
there is more often than not more than one model structure that is compatible with the data
\citep{bar_eal/15}, also see examples in \citep{fal_eal/15,ave_eal/17}. Hence there is uncertainty
not only on the model parameters, but also on the model structure. To quantify such
uncertainty remains an open problem \citep{bar_eal/15,gu_wei/18}.

\section{Parameter Estimation}
\label{s4}

By now we have chosen a model class ${\cal M}^{\rm c}$ and a specific model structure within
that class. Of course, in practice we usually choose more than a model structure to work with,
but for the sake of argument, let us focus on a single model structure. As seen in (\ref{NARX})
the model is composed by a structure (regressor variables $\psi_i$) and parameters that are 
grouped in a vector $\btheta$. Hence our model can be represented as ${\cal M}(\btheta)$ and
now the last aspect of the model that still needs determining is $\btheta$. In what follows some
of the challenges that must be faced in estimating  $\btheta$ will be discussed before mentioning
some classical estimators.

\subsection{Underlying issues}
\label{ui}

As in Sec.\,\ref{intro}, we start with a philosophical discussion which hopefully will prove helpful.
In a very loose and intuitive way, we can say that both the structure of \m and its parameters $\btheta$
should be chosen such that the estimated model \textcolor{blue}{${\cal M}(\hat{\btheta})$} be
a good representation of the system \s \hspace{-0.15cm}. The hat on $\hat{\btheta}$ indicates that
it is an estimated value of $\btheta$. 

Since we assume that at this stage
the model structure has been chosen, we move on to estimate the parameters. Hence we could
imagine a na\"ive procedure by which we chose the vector $\hat{\btheta}$ that best
approximates the model to the system, that is, $\textcolor{blue}{{\cal M}(\hat{\btheta})} \equiv$ \s \hspace{-0.15cm}.
As mentioned before, this is not a viable problem because model and system are of a very different nature.
One is a set of equations the other is a set of devices and e.g. physical or biological components, depending
on the system. Hence to move towards a solution, we measure from the system \s a set of data
$Z$, but this is still not sufficient because,  for the same reason as before,
we cannot compare a model to data, that is, to search for $\hat{\btheta}$
such that $\textcolor{blue}{{\cal M}(\hat{\btheta})} \equiv Z$  is {\it not}\, a viable problem either.
So, in order to come up with a viable problem, we produce data using the model $Z_{{\cal M}(\hat{\btheta})}$
-- which will be indicated as $Z_{\cal M}$ for short --
and now it is possible to compare data with data, because these are of the same nature. Hence,
we  search for $\hat{\btheta}$ such that $Z_{\cal M}$ is as close  to $Z$ as possible.

A standard and convenient way of measuring how close a signal is from another is to compute
the sum of squared errors:
\begin{eqnarray}
\label{J}
J = \sum_{k=1}^N  \left[ y(k)-\hat{y}(k) \right]^2 ,
\end{eqnarray}

\noindent
where $y(k)$ is the measured output and $\hat{y}(k)$ is the model output
produced with $\hat{\btheta}$. Therefore $J$ indirectly depends on $\hat{\btheta}$, because $\hat{y}(k)$
does. Because $Z=[u(k)~y(k)],~k=1,\ldots, N$ and  $Z_{\cal M}=[u(k)~\hat{y}(k)],~k=1,\ldots, N$ (notice
that $u(k)$ must be the same in both data sets), for the sake of discussion  (\ref{J}) will be
represented as $J(Z,\,Z_{\cal M})$. In this framework, if the data sets are similar then 
$J(Z,\,Z_{\cal M})$ is ``small''. Or, in other words, the smaller $J(Z,\,Z_{\cal M})$ the closer
the data sets $Z$ and $Z_{\cal M}$ are.

A remaining key issue is to decide in which way the model will be used to produce the data $Z_{\cal M}$. 
We consider two different ways of doing so by means
of a specific and simple example. Consider the model for which the parameters are assumed known
\begin{eqnarray}
\label{model}
   \hat{y}(k) & = &  [y(k-1)~u(k-1) ~ y(k-1)u(k-1)] \hat{\btheta}  ,
\end{eqnarray}
\noindent
hence the right hand side is completely known and, therefore, so is the model output $\hat{y}(k)$.
For example, the model output at time $k=10$ is given by model (\ref{model}) as:
\begin{eqnarray}   
\hat{y}(10) & = &  [y(9)~u(9) ~ y(9)u(9)] \hat{\btheta}  , \nonumber
\end{eqnarray}
\noindent
where $y(9)$ is the 9th measured sample of the output, and so on. Because we feed the
model with {\it measured}\, data and use it just to move one-step-ahead, the output computed
this way is called {\it one step ahead}\, (OSA) prediction and will be indicated by $\hat{y}_1(k)$

A different value
of $\hat{y}(10)$ would be obtained if, {\it using the same model}, we compute
\begin{eqnarray}   
\hat{y}(10) & = &  [\hat{y}(9)~u(9) ~ \hat{y}(9)u(9)] \hat{\btheta}  , \nonumber
\end{eqnarray}
\noindent
where, 
\begin{eqnarray}   
\hat{y}(9) & = &  [\hat{y}(8)~u(8) ~ \hat{y}(8)u(8)] \hat{\btheta}  , \nonumber
\end{eqnarray}

\noindent
and so on. Because this second procedure feeds the model with previously simulated
values, the output is often referred to as {\it simulated}\, or {\it free-run}\, output and will
be indicated by $\hat{y}_{\rm s}(k)$. The reader will immediately realize that the
ERR criterion is based o $\hat{y}_1(k)$ and the SRR and SMRR, on $\hat{y}_{\rm s}(k)$, see Sec.\,\ref{errsrr}.

What are the main differences between $\hat{y}_1(k)$ and $\hat{y}_{\rm s}(k)$? 
Both outputs are produced by the same model and we can think of the data revealing
or accumulating ``model signatures'' or ``model features''. Because to
produce $\hat{y}_1(k)$ the model only has to predict one step into the future starting from measured
data, the model features are somewhat hard to recognize in $\hat{y}_1(k)$. Imagine that 
the model is unstable, of course the model divergence will be in $\hat{y}_1(k)$ when
compared to $y(k)$, but -- especially if the sampling time is short -- that divergence may not
be significant. On the other hand, because $\hat{y}_{\rm s}(k)$ is produced from previously
simulated output, the model features will tend to accumulate in such data which, in a sense,
will be far more informative about the model. So, if the model is unstable, $\hat{y}_{\rm s}(k)$
will diverge to infinity.

From the previous discussion there should be no doubts that $\hat{y}_{\rm s}(k)$ is advantageous
over $\hat{y}_1(k)$ from a dynamical point of view. Hence it would be nice if we could use the
data $Z_{{\cal M}_{\rm s}}$ instead of $Z_{{\cal M}_1}$ in estimating the parameter vector. Let us
state this more formally in terms of two possible estimators that boil down to two
optimization problems:
\begin{eqnarray}
\label{prob1}
\begin{array}{lcl}
\hat{\btheta}_1 = & {\rm arg} & {\rm min}~J(Z,\,Z_{{\cal M}_1}) \\
~ & \btheta & ~\\
\end{array} 
\end{eqnarray}

\noindent
and
\begin{eqnarray}
\label{prob2}
\begin{array}{lcl}
\hat{\btheta}_{\rm s} = & {\rm arg} & {\rm min} ~J(Z,\,Z_{{\cal M}_{\rm s}}) .\\
~ & \btheta & ~\\
\end{array} 
\end{eqnarray}

For instance, the optimization problem in (\ref{prob1}) is read thus: the estimated vector $\hat{\btheta}_1$
is the argument among all possible $\btheta$ that minimizes $J(Z,\,Z_{{\cal M}_1})$, and likewise
for the problem in (\ref{prob2}).

Since we have already seen that, from a dynamical point of view, $Z_{{\cal M}_{\rm s}}$ is preferable 
to $Z_{{\cal M}_1}$, it seems natural to conclude that the estimator in (\ref{prob2}) is probably
preferable to the one in (\ref{prob1}). Well, in fact,  the estimator in (\ref{prob2}) is indeed much more
robust to noise and other aspects than  (\ref{prob1}) in many cases \citep{rib_agu/18}. This kind of reasoning
may underlie similar problems \citep{sch_eal/21}.
However, the optimization problem in (\ref{prob2})
is nonconvex, whereas the optimization problem in (\ref{prob1}) is convex and can be solved in a much
simpler way. Hence, to summarize, from a {\it dynamical }\, point of view, the solution to problem (\ref{prob2})
is preferable, but at the cost of having to solve a nonconvex optimization problem. On the other hand
to solve problem (\ref{prob1}) is {\it numerically}\, preferable, although the solution $\hat{\btheta}_1$
might not be as good and as robust as $\hat{\btheta}_{\rm s}$.

Let us close this discussion with three remarks. First, many of the ``standard'' recommendations in
system identification about testing, input persistent excitation, choice of model structure and so on
are made in order to improve the chances of $\hat{\btheta}_1$ -- which is a numerically inexpensive
solution -- being such that the model ${\cal M}(\hat{\btheta}_1)$ is dynamically acceptable
\citep{agu_eal/10}. In second
place, it must be noticed that for model classes that are {\it not}\, linear-in-the-parameters, even
problem (\ref{prob1}) is a nonconvex optimization problem. Finally, parameters of chaotic systems
can be estimated solving (\ref{prob1}) because at such short single steps, the effect of the extreme
sensitivity to initial conditions is not critical. 

\subsection{Classical estimators}

In this section we start by presenting the least squares estimator, which is the classical solution
to problem  in (\ref{prob1}), which applies to model classes that are linear-in-the-parameters. 
To see this we start considering the model structure in (\ref{NARX})
and a set of data $Z^N$. We can now write (\ref{NARX}) for any value of $k$ within the data range.
For instance, for $k=10$ we can write $y(10)=\bpsi (9)^{\T} \btheta +e(10)$ where $\bpsi (9)$ 
is the vector of regressors which goes up to time $k-1=9$ and $e(10)$ is whatever cannot be
explained in the data at time $k=10$ using $\bpsi (9)^{\T} \btheta$. Hence we would like to find $\btheta$ in such a way as
to minimize the unexplained part $e(10)$. In order to gain robustness, we consider the model
structure (\ref{NARX}) along all the data, which we now call the {\it identification data set}\, $Z^N$.
That means that for each value of $k$ within the data range we will have an equation of the 
form $y(k)=\bpsi (k-1)^{\T} \btheta +e(k)$ which can be expressed as a matrix equation thus:
\begin{equation}
\label{eqn_errorm}
	\by  =  \Psi \btheta + \be,
\end{equation}

\noindent
where $\Psi \in \bbR^{N \times n_\theta}$,\footnote{In fact, because of the lags in 
the model, a data set of size $N$ will result in slightly smaller set of equations. We will
not detail this to keep nomenclature as light as possible.}
$\btheta \in \bbR^{n_{\theta}}$ and $\by\! = \! \{y(k)\}_{k=1}^N$. 
Suppose we have an estimate of the parameter vector $\hat{\btheta}$.
Then (\ref{eqn_errorm}) can be rewritten as: 
\begin{eqnarray}
\label{eqn_errorm2}
	\by  & =  & \Psi \hat{\btheta} + \bxi, \\
	& = & \hat{\by}_1 + \bxi \nonumber
\end{eqnarray}

\noindent
where $\bxi \! = \! \{\xi(k)\}_{k=1}^N$ is the vector of {\it residuals}\, which can be taken
as an estimate of the unknown noise $\be$ under ``favorable'' conditions and 
$\hat{\by}_1$ is the vector of OSA predictions.

One way of determining $\hat{\btheta}$ if $\Psi$ does not contain columns correlated
with $\bxi$ (as in the case of NARX models)
is by the Least-Squares (LS) estimator, that minimizes the mean squared value
of $\bxi = \by - \hat{\by}_1$  and 
is given by 
\begin{equation}
\label{LS}
\hat{\btheta}_{\rm LS}=(\Psi^{\T}\Psi)^{-1}\Psi^{\T}\by ,
\end{equation}

\noindent
and is the solution to problem (\ref{prob1}).

In the classical LS solution (\ref{LS}) all the residuals receive equal weight, that is
$\xi(k) = y(k) - \hat{y}_1(k),~\forall k$. There might be reasons in some situations to
give specific weights to the residuals as $w(k)\xi(k)$. Placing $w(k),~k=1,\,\ldots , N$ along the 
diagonal of a matrix $W \in \bbR^{N\times N}$, the solution
\begin{equation}
\label{WLS}
\hat{\btheta}_{\rm WLS}=(\Psi^{\T}W\Psi)^{-1}\Psi^{\T}W\by 
\end{equation}

\noindent
is known as the weighted least squares estimator (WLS). There are two classical choices for $W$.
If the weighting matrix is taken as the covariance matrix for the noise, the WLS becomes
the so called Markov estimator which is akin to the generalized least squares estimator. A
second important choice of $W$ is to take the weight of the present moment as $w(k)=1$, 
the previous one as $w(k-1)=\lambda$, then $w(k-2)=\lambda^2$ and so on, where $\lambda<1$
is known as the forgetting factor. Hence the WLS estimator is useful to derive a recursive
LS estimator with forgetting factor.

The optimization problems in (\ref{prob1}) and (\ref{prob2}) are said to be unconstrained.
This means that all the degrees of freedom are used to minimize the functional $J$, and
the solution can be any vector of real values $\hat{\btheta} \in \bbR^{n_\theta}$. Now suppose
that there is a set of constraints on the parameter vector written as ${\bf c}=S{\btheta}$,
where ${\bf c}$ is a given constant vector, and $S$ is a known constant matrix. This means
that no matter what parameter vector is chosen to minimize $J$, it must simultaneously
satisfy the set of constraints ${\bf c}=S{\btheta}$. Hence, there are less degrees of freedom
available to minimize $J$. The solution to the problem 
\begin{eqnarray}
\label{prob3}
\begin{array}{lcl}
\hat{\btheta}_{\rm CLS} = & {\rm arg}~ & ~{\rm min}~J(Z,\,Z_{{\cal M}_1}) \\
~ & \btheta: {\rm c}=S\btheta , & ~\\
\end{array} 
\end{eqnarray}

\noindent
is given by \citep{dra_smi/98}
\begin{equation}
\label{constrained}
 \hat{\btheta}_{\rm CLS} = (\Psi^{\T}\Psi)^{-1}\Psi^{\T}\by-
        (\Psi^{\T}\Psi)^{-1}S^{\T}[S(\Psi^{\T}\Psi)^{-1}S^{\T}]^{-1}
        (S\hat{\btheta}_{\rm LS}-{\bf c}).
\end{equation}

As it will be shown in future sections, the solution (\ref{constrained}) turns out to be very
helpful when we are able to translate auxiliary information about the system \s in terms
of a set of equality constraints on the parameters. Of course, this is easier to do for some
model classes and hence this serves as motivation for choosing certain model classes
depending on the available auxiliary information and our ability to code it in terms of constraints.

The issue of parameter estimation -- or model training as known in other fields -- is vast. A key-point
which has not been dealt with because it would require a far more technical discussion has to
do with the statistical properties of the noise $\be$ in (\ref{eqn_errorm}). Here it has been
implicitly assumed that it is white. Whenever the noise is correlated, the LS estimator becomes
biased. There are several solutions to this problem, for instance: i)~if one can find a
set of variables, called instruments, that are not correlated to the noise and remain correlated
to the output, the {\it instrumental variable}\, estimator can be used \citep{you/70}, ii)~the model
class can be changed to try to model the correlation in the noise. A common choice is to 
add a {\it moving average}\, (MA) part to the model. The resulting model class is no longer linear-in-the-parameters
and iterative estimators are available such as the {\it extended least squares}\, \citep{che_bil/89,lu_eal/01},
and iii)~by solving the problem in (\ref{prob2}) \citep{agu_eal/10}.

The basic theory and algorithms can be found in most text-books some of which will be 
mentioned in Sec.\,\ref{fr}. There are many alternative procedures for polynomial models
such as \citep{nep_eal/07,wei_bil/09} and also for other representations such as
rational \citep{wu_eal/08,zhu_bil/93}, neural networks \citep{che_eal/90nn,mas_eal/93} 
and radial basis functions \citep{che_eal/90rbf}.

We close mentioning that although technical aspects of the estimators certainly depend on
the model class used, many discussions in this section, especially those in Sec.\,\ref{ui}
remain valid for other model classes \citep{rib_agu/18}.

\subsection{The danger of overparametrization}

From what has been discussed in this section it is now possible to have a better understanding of
the danger of overparametrizing a model. Let us consider problem (\ref{prob1}) again. Would it
be possible to find $\hat{\btheta}$ such that $J(Z,\,Z_{{\cal M}_1})=0$? It might sound absurd
(and in practice it is, in fact), but we can force $\hat{y}_1(k)=y(k),~\forall k$ {\it for the 
estimation data}. Hence, let us see under which conditions $\bxi=0$ in (\ref{eqn_errorm2}). For this, lets go back 
one step and consider (\ref{eqn_errorm}). If we increase the size of $\btheta$ to the point that
$n_\theta =N$, then $\Psi \in \bbR^{N \times N}$ becomes a square matrix. Assuming that it is
nonsingular, then $\hat{\btheta}=\Psi^{-1}\by$ and therefore $\bxi=0$. What we have done is
to increase the number of degrees of freedom (parameters to be estimated) to match the number of 
``constraints'' (the number of rows of $\Psi $) to end up with a square system of equations that
has a single solution. Is that solution any good? Most likely not. And the reason is that all the
uncertainties and noise in the data $y(k)$ will be fit by the model, which ideally should only fit
the underlying dynamics of the system. The resulting model in this hypothetical case has a 
lot more parameters than needed -- remember that $n_\theta =N$ -- and therefore 
the model is said to be overparametrized.

An interesting remark on which we will not dwell here is this: if we consider  problem (\ref{prob2})
instead, we will not face the same problem. The reason is that in order to have $J(Z,\,Z_{{\cal M}_{\rm s}})=0$
we would need to have $\hat{y}_{\rm s}(k)=y(k),~\forall k$. This will not be achieved in practice
because $\hat{y}_{\rm s}(k)$ is obtained by iterating the model (see Sec.\,\ref{ui}) and an
overparametrized model will, more often than not, be unstable, hence $\hat{y}_{\rm s}(k) \neq y(k)$.
This discussion then has two messages. First, problem (\ref{prob2})
is definitely more robust than problem (\ref{prob1}). Second, if the latter problem is 
the one being solved, then overparametrization is a real danger especially for 
NARMAX polynomial models and similar model classes.

\section{Model Validation}
\label{s5}

We finally get to the last step in system identification. Here we would like to answer the question,
does the model ${\cal M}(\hat{\btheta})$ represent the underlying dynamics of the system?

One of the basic rules in model validation is to use a set of data $Z_{\rm v}$ for this purpose which
is different from the data used to fit the model, that is the estimation or validation data $Z$. This is necessary because we cannot compare
the model directly to the system \s \hspace{-0.15cm} as mentioned before.
Then, in order to compare entities that are alike we take data from the system and data from the model
$Z_{\cal M}$ and compare both using some kind of metric $V(Z_{\rm v}, \, Z_{\cal M})$. It should come 
as no surprise that the choice of $Z_{\rm v}$, $Z_{\cal M}$ and $V$ have their influence on the validation.
If there is only one set of data $Z^N$, then this should be divided into a part for identification 
$Z^{N_{\rm i}}$ and a part for validation $Z_{\rm v}^{N_{\rm v}}$, such that $N=N_{\rm i}+N_{\rm v}$.
A common choice is to split the data such that $N_{\rm i}= 2N_{\rm v}$, but this is not a must.

As for  $Z_{\rm v}$, it is recommended that these data be measured from the system operating 
over a range (amplitude and frequency-wise) consistent with the intended use of the model. In
obtaining the identification data $Z$ it is reasonable to try to excite the system over the widest possible
operating range (in the nonlinear case) in order to gather as much information as possible. When
it comes to validation, there is no reason why we should request that the model be a faithful representation
of the system over operating ranges that will not be visited in practice.

Concerning the choice of $Z_{\cal M}$, as discussed in Sec.\,\ref{ui}, the use of $Z_{{\cal M}_{\rm s}}$
should be strongly preferred over $Z_{{\cal M}_1}$. As a matter of fact, the latter is close to useless
for model validation, because even poor models can produce OSA predictions that are close to the
validation data. On the other hand, the use of  $Z_{{\cal M}_{\rm s}}$ is very exacting, and that is what
is required in validation.

Finally, the use of a chosen metric $V$ should be considered with care. Suppose that $V$ is a
measure of distance between two data sets. Also, consider two models with
$Z_{{\cal M}_{\rm s1}}$ and $Z_{{\cal M}_{\rm s2}}$ such that
$V(Z_{\rm v}, \, Z_{{\cal M}_{\rm s1}})< V(Z_{\rm v}, \, Z_{{\cal M}_{\rm s2}})$. This only means that
the first model is better than the second {\it with respect to the chosen metric V}. A certain metric
might not be sufficiently specific to capture the dynamical aspects of the models. Common
metrics are the mean squared error or the mean absolute percentage error, but the list is much longer.

Before mentioning some specific criteria, let it be said that the question: ``is this model valid?'' should be
preceded by the question: ``what is the intended use for the model?'' In other words, the validity
of a model should not be tackled as an absolute attribute, but rather as something that should make
sense within a context. That is perhaps one of the greatest difficulties in model validation because
very often we do not know exactly what to ask from the model.

\subsection{Residual tests}

This consist of testing the residual vector $\bxi$ for whiteness and for correlations with the 
input and output \citep{bil_tao/91,bil_zhu/94b}. The main idea behind
this procedure is that if the model structure is sufficiently flexible and the estimation
procedure was sound, then all the correlations in the data should be explained by the model,
hence the residuals should not be self-correlated and uncorrelated with the input and output.
Also, because the standard linear correlation functions are unable to detect all nonlinear
correlations, then specific nonlinear correlation functions should be used
\citep{bil_zhu/95,zhu_eal/07}.

Because the residuals are defined based on the OSA prediction, clearly a model that
passes the residual tests might still be inadequate from a {\it dynamical}\, point of view.
Also, residual tests are not  sensitive to overparametrization. That is, overparametrized
models often pass residual tests.

\subsection{Dynamical invariants}

A rather particular situation arises when the system \s is chaotic. In that case it is quite
common to use a number of dynamical and geometrical invariants to quantify the dynamics. 
Among such indices one can mention correlation dimension, Lyapunov exponents, Poincar\'e
sections and bifurcation diagrams. Some of theses tools have been discussed in the
context of model validation in \citep{hay_bil/94,agu_bil/93a}.

A more exacting procedure for model validation, but now restricted to 3D chaotic models is
topological analysis \citep{let_eal/95b,let_eal/96b,let_eal/00soofi}. This procedure
builds topological templates from the system and model data and compares them.
From such templates it is possible to extract the population of periodic orbits and to compare
the model population the that of the original system. This is far more detailed, intricate and
complicated than to just compare the appearance of attractors.

\subsection{Synchronization}

A very interesting procedure suggested in \citep{bro_eal/94pre} is that of synchronization.
The idea is that that underlies the concept of a state observer. An error signal between model output and data is fed back
 to the model to force the model output to follow the data. From the days of Huygens it is
 known that two coupled oscillators will synchronize if their natural frequencies (their intrinsic
 dynamics) be close. The same applies here. We would like to know if the model
 dynamics is close to the dynamics underlying the data. Hence we couple model and
 data and see if the ``synchronization error'' becomes small.
 
 This procedure has two practical problems. First, in many cases synchronization can be
 achieved at the expense of increasing the coupling force. It has been pointed out that
 different dynamics can be made to synchronize and therefore synchronization has little
 to offer when it comes to (absolute) model validation \citep{agu_eal/06pre,let_eal/18}. 
 Second, if the model does synchronize with large coupling, does this imply that the 
 underlying dynamics have been learned by the model? The answer to this question is negative.
 However, if the model and data have the same dynamics, the coupling required for
 synchronization can be very small.
 
 A practical approach to a similar problem has been put forward in \citep{agu_eal/06pre}.
 There the main goal is not to validate in absolute terms a model, but rather compare
 models and choose the one that is closest to the dynamics underlying the data. This was
 referred to as {\it model evaluation}\, in the aforementioned reference. To achieve this
 a measure of synchronization effort must be taken into account for the comparison to make
 any reasonable sense. 
Then if the maximum synchronization error is very different, it can be used to rank the models:
the best model has the smallest maximum error. However, often models have similar errors,
in that case the synchronization effort is used. Hence given two models with similar errors,
the one that required a smaller effort is the better model. So, in short synchronization
allied to a measure of cost can help in comparing models, but not to get any general
assessment of model quality.

\section{Grey-Box Techniques}
\label{gb}

So far we have described system identification problems from a black-box point of view, that
is, the only source of information about the system \s is the data set $Z$. Even if we assume
that, say, the model order ($n_y$) and degree of nonlinearity ($\ell$) are known, still we refer
to this as black-box identification.

\begin{figure}[!ht]
		\centering
		\includegraphics[width=0.80\textwidth]{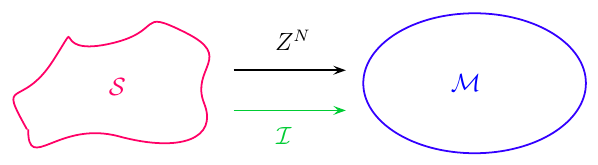}
		\caption{\small Simplified schematic diagram for grey-box identification, where \s represents the system
		that should be approximated by a model \m which is built from a set of
		measured data $Z^N$ of length $N$ {\it and}\, auxiliary information~\I \hspace{-0.15cm}.}
		\label{f110219a}
\end{figure}

Now we move on to grey-box system identification problems where it is assumed that some
additional information is available, which we will represent by \I as in Figure~\ref{f110219a}.
This raises a number of interesting questions, such as:

\begin{enumerate}
\item what kind of auxiliary information \I is useful?
\item how does \I relate to the model class ${\cal M}^{\rm c}$?
\item assuming that \I and \m$\in {\cal M}^{\rm c}$ are compatible, how do we actually use
\I in determining $\textcolor{blue}{{\cal M}(\hat{\btheta})}$?
\end{enumerate}

In what follows a glimpse to some grey-box techniques will be presented, and literature will be
provided for details.

\subsection{\I is the static function}

We now assume that \I is the static function (calibration curve) of the system.
To see this we must first consider the steady-state analysis of a NARX model.

Consider the NARX and deterministic part of model (\ref{NARMAX}) taken in steady-state, that is,
$\bar{y}=y(k-i),~\forall i$ and $\bar{u}=u(k-j),~\forall j$, then we have an {\it algebraic}\, polynomial
of the form
\begin{eqnarray}
\label{NARXss}
   \bar{y} & = & F_{\rm ss}^\ell [ \bar{y},\, \bar{u} ]  .
\end{eqnarray}

\noindent
It is left to the reader to verify that the parameters of the right-hand-side of (\ref{NARXss}) are
the cluster coefficients \citep{agu_bil/94b,agu_eal/02iee}. Equation (\ref{NARXss}) says that if
a constant input $\bar{u}$ is applied to the NARX dynamical model, then in steady-sate, if
the model is asymptotically stable, the output $y(k)$ will settle to $\bar{y}$. This is very closely related to the
concept of fixed-points of autonomous models  and the number of such
equilibria is {\it given by the highest power of the output}, for which $\ell$ is an upper bound \citep{agu_men/96}.
For example, model
\begin{eqnarray}
   y(k) = \theta_1 y(k-1) + \theta_2 y(k-1)^3 + \theta_3 u(k-2)y(k-1) \nonumber
\end{eqnarray}

\noindent
has three fixed points for each constant value of the input, whereas the model
\begin{eqnarray}
   y(k) = \theta_1 y(k-1) + \theta_2 y(k-1)^2u(k-1) + \theta_3 u(k-2)y(k-1), \nonumber
\end{eqnarray}

\noindent
for which $\ell=3$ -- as for the previous one -- only has two fixed points for each constant value of the input.

In cases for which there is a ``calibration curve'' for the system this will correspond
to $F_{\rm ss}^\ell$ in (\ref{NARXss})  and no output regressors of degree 2 or higher are needed
to reproduce the calibration curve, as seen in the next example.

\begin{figure}[htb] 
\centering
\includegraphics[scale=0.4]{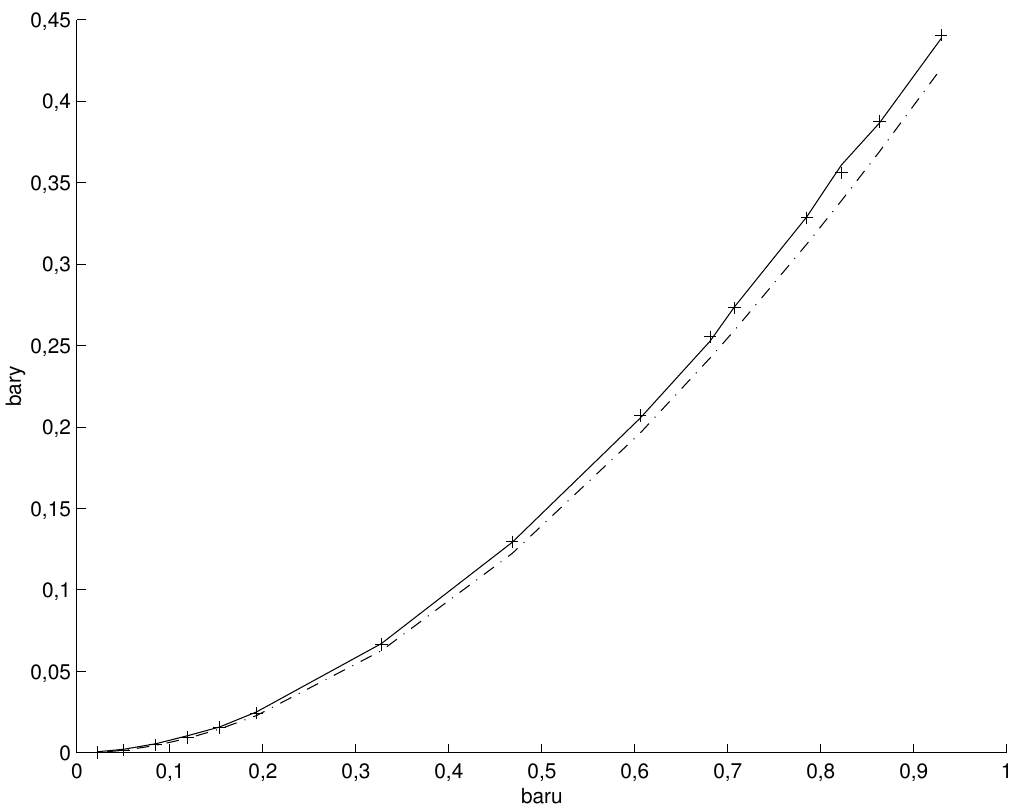}
\caption{\small \label{cassini}Static functions of identified models 
(---)~measured;  \hbox{(-$\cdot$-)}~unconstrained model
(\protect{\ref{modelo cassini cp MQ}}) and  \hbox{($\times$)}~constrained model
(\protect{\ref{modelo cassini cp MQR}}).}
\end{figure}

\paragraph{Example 1.} This example uses data measured from a small thermal pilot plant
\citep{agu_eal/02iee}. Because there is no multi-stability, that is, for a given input in
steady-state $\bar{u}$ there will be only {\it one}\, value for the output $\bar{y}$, it is possible to
remove all potential regressors from clusters $\Omega_{y^p},~p>1$. That is, quadratic and cubic
regressors in $y$ need not be considered, nor crossed terms  $\Omega_{y^pu^m},~p>1,~\forall m$.
Besides, because the static data look quadratic (see Fig.\ref{cassini}), the cluster $\Omega_{u^3}$ was
also removed from the set of candidate regressors. This set of actions is already due to what we
find in \I \hspace{-0.15cm}. 

The following black-box model was then obtained:
\begin{eqnarray}
\label{modelo cassini cp MQ}
y(k) &=& 1.2929y(k-1) + 0.0101u(k-2)u(k-1) + 0.0407u(k-1)^{2} \nonumber \\
     & & - 0.3779y(k-2) -0.1280u(k-2)y(k-1) \nonumber \\
     & & + 0.0957u(k-2)y(k-2) + 0.0051u(k-2)^{2}.
\end{eqnarray}

\noindent
In steady state, model (\ref{modelo cassini cp MQ}) can be written thus
\begin{eqnarray}
\bar{y} &=& 1.2929\bar{y} + 0.0101\bar{u}^2 + 0.0407\bar{u}^2  - 0.3779\bar{y} -0.1280\bar{u}\bar{y} \nonumber \\
     & & + 0.0957\bar{u}\bar{y} + 0.0051\bar{u}^2,
\end{eqnarray}

\noindent
which has a static nonlinearity of the form
\begin{equation}
\label{sf1}
\bar{y} = \frac{\Sigma_{u^2}\bar{u}^{2}}{1 - \Sigma_y - \Sigma_{yu} \bar{u}}.
\end{equation}

\noindent
In a typical grey-box problem, 
we want to estimate the cluster coefficients $\Sigma_{u^{2}}~ \Sigma_{yu} ~\Sigma_{y}$ in
such a way that (\ref{sf1}) fits the measured static data Fig.\,\ref{cassini}. Because, (\ref{sf1})
is not linear-in-the-parameters, then such a fit must be accomplished by some nonconvex optimization
algorithm. Here a quasi-Newton FBGS method was used to find: 
$\textbf{c} = [\Sigma_{u^{2}}~ \Sigma_{yu} ~\Sigma_{y}]^{\T} = [0,0615~ -0,0360~~ 0,9128]^{\T}$.
This can then be written in terms of a set of three linear constraints

\begin{eqnarray}
\label{rest}
\left[\begin{array}{c}
      0,0615 \\
     -0,0360 \\
      0,9128
      \end{array}
\right] & = & 
\left[\begin{array}{ccccccc}
0& 1& 1& 0& 0& 0& 1 \\
0& 0& 0& 0& 1& 1& 0 \\
1& 0& 0& 1& 0& 0& 0
\end{array}
\right]
\left[\begin{array}{c}
     \theta_{1} \\
     \theta_{2} \\
     \theta_{3} \\
     \theta_{4} \\
     \theta_{5} \\
     \theta_{6} \\
     \theta_{7}
      \end{array}
\right]. \\
{\bf c} & = & S \btheta . \nonumber
\end{eqnarray}

\noindent
Notice that $\theta_2$, $\theta_3$ and $\theta_7$ are the only parameters of the term cluster $\Omega_{u^2}$ (quadratic
input terms), hence their addition is $\Sigma_{u^2}$ by definition. That composes the first constraint in
(\ref{rest}) and so on. 

The set of constraints in (\ref{rest}) and the parameter vector $\hat{\btheta}_{\rm LS}$ of the model in (\ref{modelo cassini cp MQ}) --
which was obtained by unconstrained estimation -- can be used with the constrained estimator (\ref{constrained})
to yield the model (notice that the model structure is the same):
\begin{eqnarray}
\label{modelo cassini cp MQR} 
y(k) &=& 1.2796y(k-1) + 0.0178u(k-2)u(k-1) +
    0.0408u(k-1)^{2} \nonumber \\
     & &  -0.3668y(k-2) -0.2565u(k-2)y(k-1) \nonumber \\ 
     & & + 0.2205u(k-2)y(k-2) + 0.0029u(k-2)^{2},
\end{eqnarray}

\noindent
which {\it by construction}\, has the desired static function.

As will be illustrated below in Example~2, there is a simpler way to proceed which will not require 
a non convex optimization step. The price paid will be that the full static function will
not be imposed, but rather only some user-chosen steady-state data points.

In closing, it is interesting to point out that \I can be used either way. Here, 
we assume that \I is known beforehand and we use such info in building the model.
Moving in the other direction, it is possible to, from an estimated model, to
extract, say, the static function \citep{agu/97pd,agu_eal/02iee}. This last procedure
proves to be very handy in estimating Hammerstein and Wiener models from 
identified NARX models \citep{agu_eal/05}.

\subsection{\I is steady-state data}
\label{subssd}

The next case that will be considered is when the auxiliary information \I is a set of
steady-state data
\hbox{$\bar{\bf u}=[ \bar{u}_1, \ldots, \bar{u}_M ]^{\T}$} and the corresponding
output values $\bar{\bf y}=[ \bar{y}_1, \ldots, \bar{y}_M ]^{\T}$. 
Let us indicate the steady-state data as \I$=Z_{\rm ss}^M=[\bar{\bf u}, \bar{\bf y}]$.
Of course, we still assume that there is the set of dynamical data
$Z^N= [ u(k), y(k) ],~k=1,\,2,\ldots , N$. A loose grey-box problem would be: given $Z^N$ and 
$Z_{\rm ss}^M$, find a model $\textcolor{blue}{{\cal M}(\hat{\btheta})}$ that {\it simultaneously}\,
uses both data sets. In what follows this aim will be made more precise.

If the dynamics of the system \s do not change significantly over a certain range of
operating conditions, but only the gain, that is, the static function is nonlinear, then
the procedure to be described in what follows is a workable solution to not having to
perform large-amplitude dynamical tests on the system. The dynamical data $Z^N$
can be collected over a rather narrow operating range and the information of the
system static nonlinearity {\it far}\, from such a range comes in through $Z_{\rm ss}^M$.

\subsubsection{Constraining parameters}
\label{sscp}

One way of using \I$=Z_{\rm ss}^M=[\bar{\bf u}, \bar{\bf y}]$ in addition to $Z^N$ is to use
\I to derive parameter constraints and then use such result with (\ref{constrained}). This was
already done in Example~1, but there the set of constraints was built from a nonlinear fit
of an algebraic relation (\ref{sf1}) to the static data. Here we want to avoid solving a nonconvex 
optimization problem and to take the constraints directly from the data. To see how to do this,
notice that in the case of NARX polynomials the steady-state relation (\ref{NARXss}) will
yield a linear constraint for each steady-state point in $Z_{\rm ss}^M=[\bar{\bf u}, \bar{\bf y}]$.
Such a constraint can be then used in (\ref{constrained}) which also uses $Z^N$. It should be
noticed that not all model classes will yield linear constraints. Therefore there {\it is}\, a connection
between the model class and \I \hspace{-0.15cm}. This will be seen in the following example.

\paragraph{Example 2.} Let us start with a simple hypothetical example. Let us assume that
the model structure is given by
\begin{eqnarray}
\label{modelX}
y(k) & = & \theta_1 y(k-1)+ \theta_2 y(k-2) + \theta_3 u(k-1) + \theta_4 u(k-2)^2 \nonumber \\
 & ~& + \theta_5 u(k-1)u(k-2) + \theta_6 u(k-2) .
\end{eqnarray} 

It is further assumed that we have dynamical and static data $Z$ and $Z_{\rm ss}$. To estimate
the unknown parameters $\theta_i,~i=1,\, \ldots, 6$ from $Z$ is well-known problem. For instance,
using the classical Least Squares estimator (\ref{LS}) the parameter vector $\hat{\btheta}_{\rm LS}$
is readily obtained. Now suppose that
apart from using $Z$ we want to make sure that the static function of (\ref{modelX}) passes through
two of the points in $Z_{\rm ss}$, say $[\bar{u}_3,\,\bar{y}_3]$ and $[\bar{u}_7,\,\bar{y}_7]$. 
We start by writing model (\ref{modelX}) in steady-state thus:
\begin{eqnarray}
\bar{y} =  (\theta_1 + \theta_2)\bar{y} + (\theta_3 +\theta_6) \bar{u} + (\theta_4 + \theta_5)\bar{u}^2 \nonumber .
\end{eqnarray} 

\noindent
Hence the two constraints are
\begin{eqnarray}
\bar{y}_3 & = &  (\theta_1 + \theta_2)\bar{y}_3 + (\theta_3 +\theta_6) \bar{u}_3 + (\theta_4 + \theta_5)\bar{u}_3^2 \nonumber \\
\bar{y}_7 & = &  (\theta_1 + \theta_2)\bar{y}_7 + (\theta_3 +\theta_6) \bar{u}_7 + (\theta_4 + \theta_5)\bar{u}_7^2 ,\nonumber 
\end{eqnarray} 

\noindent
which can be rewritten as  ${\bf c}=S{\btheta}$ with
\begin{equation}
{\bf c}=
\left[\begin{array}{c}
     \bar{y}_3\\
     \bar{y}_7
 \end{array}
\right] ;
\hspace{0.5cm}
S=
\left[\begin{array}{ccccccc}
\bar{y}_3& \bar{y}_3& \bar{u}_3 & \bar{u}_3^2& \bar{u}_3^2& \bar{u}_3 \\
\bar{y}_7& \bar{y}_7& \bar{u}_7 & \bar{u}_7^2& \bar{u}_7^2& \bar{u}_7 
\end{array}
\right]. \nonumber
\end{equation}

\noindent
Now, as in Example~1, with ${\bf c}$, $S$ and $\hat{\btheta}_{\rm LS}$, 
the constrained estimator (\ref{constrained}) can be used to find another parameter
vector  $\hat{\btheta}_{\rm CLS}$ which {\it exactly}\, satisfies the two aforementioned constraints
and {\it approximately} -- in a least squares sense -- fits the dynamical data $Z$.

%

%

\subsubsection{Imposing a transcritical bifurcation}

The reason for inserting an example on the imposition of a fixed-point bifurcation
here demands a short explanation that can be announced in two steps. First, the 
normal equations of fixed-point bifurcations for maps are (first-order) difference equations.
Second, bifurcations relate to steady-state behavior of dynamical systems. Therefore
the conditions for the occurrence of certain bifurcations can be found from 
first-order autoregressive equations. In a sense, a bifurcation diagram can be seen as
a particular case of a steady-state function of a nonlinear polynomial model if the
input is used as ``bifurcation parameter''. The following example illustrates the case
for a transcritical bifurcation which has the appearance of a static function with
dead-zone. Other examples include period-doubling
bifurcation \citep{agu_fur/07} and the Hopf (Neimark-Sacker) bifurcation \citep{agu_rod/12}.

\paragraph{Example 3.} For details on this example see \citep{agu/14}.
Consider a Hammerstein system with static function given by:
\begin{eqnarray}
z(k) = \left\{
\begin{array}{lc}
0, & {\rm if}~ u(k)<1 \\
u(k)-1, & {\rm if}~  u(k) \ge 1,
\end{array}
\right. \nonumber
\end{eqnarray}

\noindent
with linear dynamics described by:
\begin{equation}
\label{a200614}
y(k)=0.9y(k-1)+0.7z(k-1)+e(k),
\end{equation}

\noindent
where the input is 
$u(k)\sim {\rm U}(1,\,0.6)$, that is a uniform random signal with mean equal to one and
standard deviation $\sigma=0.6$. The output was simulated using
(\ref{a200614}) with $e(k) \sim {\rm WGN}(0,\,0.1)$ which is a zero mean white Gaussian noise 
with standard deviation  $\sigma=0.1$. The data
$Z=[u(k)~y(k)],~k=101,\ldots 300$ were used to estimate three models. The model structure was
automatically determined using the ERR criterion (\ref{eq_errp}). The models were tested on the validation 
data $Z_{\rm v}=[u(k)~y(k)],~k=301,\ldots 500$.

\begin{figure}[!ht]
\centering
\begin{tabular}{c}
\includegraphics[scale=0.4]{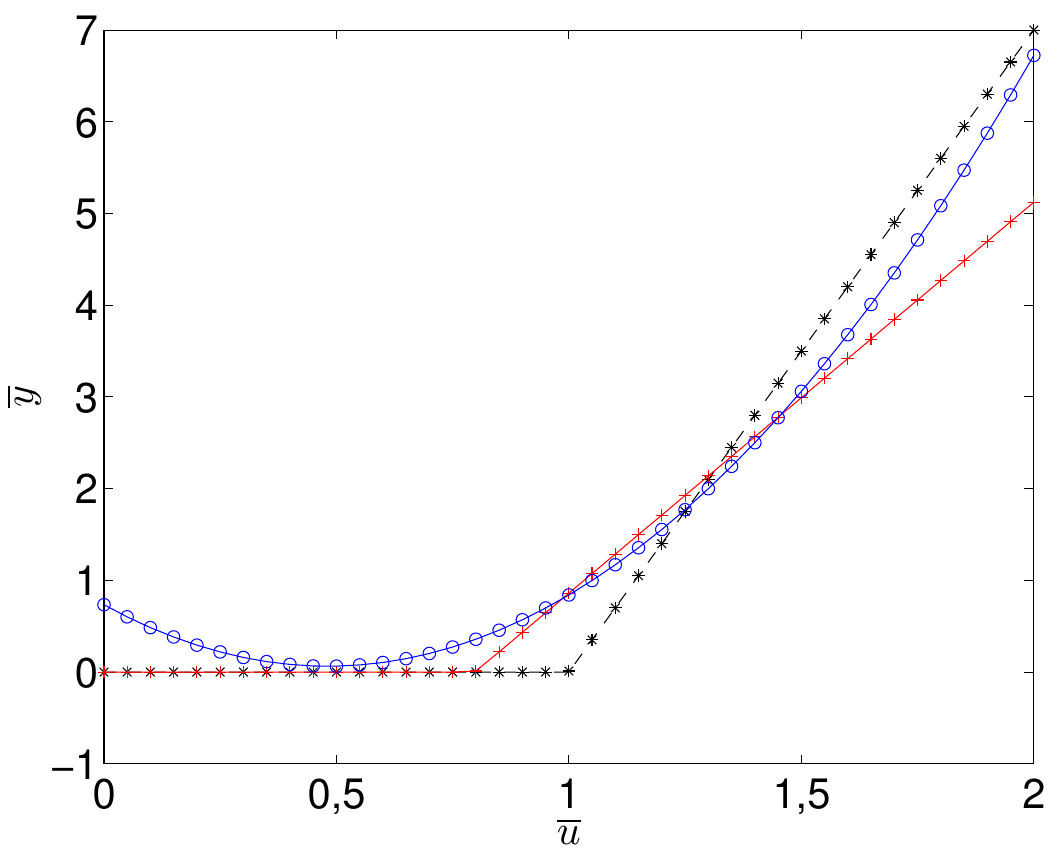}
\end{tabular}
\caption{\small \label{fig:dg25t}Black dashed is the original nonlinear static function;
blue circles indicate the static function of black-box model  ${\mathcal M}_1$ (\ref{b200614}); the red crosses
correspond to grey-box model ${\mathcal M}_2$ (\ref{model3})
and the asterisks, to another grey-box model ${\mathcal M}_3$ (\ref{d200614}).}
\end{figure}

The black-box model ${\mathcal M}_1$ was:
\begin{eqnarray}
\label{b200614}
y(k)& = & +0.8676  \, y(k-1)  +0.37369  \, u(k-1)^2 -0.36828  \, u(k-1)  \nonumber \\
        &   &  +0.09723 +0.86609 \! \times \! 10^{-2} \, u(k-3)^2 ,
\end{eqnarray}

\noindent
for which the static behaviour is shown in Figure~\ref{fig:dg25t} which is clearly rather poor. 
This would be expected because, being a NARX polynomial (\ref{b200614}) also has a
polynomial and therefore smooth static function, as indicated in Figure~\ref{fig:dg25t} by
the blue tracing.

We move a step forward and assume that the static function is formed by two line
segments. In order to make this more understandable, the following result is quoted from \citep{agu/14}:
\begin{lemma}
\label{lemma1}
A NARX polynomial model with $\ell \ge 2$ for which the only nonzero cluster coefficients are
$\Sigma_y$, $\Sigma_{uy}$ and $\Sigma_{y^2}$ has the following two
sets of equilibria:
\begin{eqnarray}
\label{fp}
\bar{Y}_1 = \bar{y}=0 \hspace{0.5cm}{\rm and} \hspace{0.5cm}
\bar{Y}_2(\bar{u}) & = & \bar{y}= \frac{1-\Sigma_y}{\Sigma_{y^2}} - \frac{\Sigma_{uy}}{\Sigma_{y^2}} \bar{u} , \nonumber
\end{eqnarray}
which intercept at
\begin{equation}
\label{bifpoint}
\bar{u}_{\rm c}=\frac{1-\Sigma_y}{\Sigma_{uy}}. \nonumber
\end{equation}
\end{lemma}

\noindent
If, in addition, the sets of equilibria $\bar{Y}_1$ and $\bar{Y}_2$ exchange stability at $\bar{u}_{\rm c}$ we
have a {\it transcritical bifurcation}.

The information in Lemma~\ref{lemma1} can be used -- see \citep{agu/14} for details -- to obtain
a model ${\mathcal M}_2$ by removing $\Omega_u$ and the constant term from the set of
candidates:

\begin{eqnarray}
\label{model3}
y(k) & = & +0.80721 \, y(k\!-\!1) +0.24200  \, u(k\!-\!1)^2 \nonumber \\  
        &   & -0.10572 \, u(k\!-\!3)u(k\!-\!1) \!+\!0.02141 \, u(k\!-\!3)y(k\!-\!1) \nonumber \\
        &   & +0.03838 \, u(k\!-\!3)^2+0.63554 \times 10^{-2} \, y(k\!-\!1)^2 ,
\end{eqnarray}

\noindent
which is also unable to fit the static function of the system, although the auxiliary information
that the static function is composed of two line segments (these are the two sets of equilibria
mentioned in Lemma~\ref{lemma1}) was correctly implemented, 
as seen in Figure~\ref{fig:dg25t}.

We now use the information in Lemma~\ref{lemma1} which relates to the location of the break point at $\bar{u}_{\rm c}=1$ 
and the inclination $\alpha=7$ of the second part of the static function ($\bar{Y}_2(\bar{u})$ in the lemma). Hence
the following constraints can be written
\begin{eqnarray}
\label{constraintsab}
0  =  \Sigma_{u^2}, \hspace{0.8cm}
1  =  \frac{1-\Sigma_y}{\Sigma_{uy}}, \hspace{0.8cm}
7  = -\frac{\Sigma_{uy}}{\Sigma_{y^2}}.
\end{eqnarray}

\noindent
The second, third and fifth parameters of (\ref{model3}) compose $\Sigma_{u^2}$. Therefore the
summation of the three of them must be zero. This is the first constraint. The other two are built
in a similar way. Hence the set of constraints is in the form ${\bf c}=S{\btheta}$ with
\begin{equation}
{\bf c}=
\left[\begin{array}{c}
     0 \\
     1 \\
     0 
      \end{array}
\right] ;
\hspace{0.5cm}
S=
\left[\begin{array}{ccccccc}
0& 1& 1& 0& 1& 0 \\
1& 0& 0& 1& 0& 0 \\
0& 0& 0& 1& 0& 7
\end{array}
\right]. \nonumber
\end{equation}

\noindent
Finally, model ${\mathcal M}_3$ was estimated using  (\ref{constrained})
\begin{eqnarray}
\label{d200614}
y(k)& = & +0.82469  \, y(k-1) +0.25589  \, u(k-1)^2 \nonumber \\
         &   & -0.15788  \, u(k\!-\!3)u(k\!-\!1) \!+\!0.17531\, u(k\!-\!3)y(k\!-\!1) \nonumber \\
         &   & -0.09801  \, u(k-3)^2 -0.02505 \, y(k-1)^2   ,
\end{eqnarray}

\noindent
which is able to reproduce the dead-zone type of static function (see Figure~\ref{fig:dg25t}). 
Once the auxiliary information \I is described in terms of constraints (\ref{constraintsab}), the
constrained least squares estimator  (\ref{constrained}) can be readily used. The trick then is to be able to 
translate  \I into constraints that, of course, will depend on the model class being used.
The details of that development for this and similar examples can be found in 
\citep{agu/14}.

It is noted that the root mean square error on the dynamical data is 0.222 for ${\cal M}_1$,
0.350 for ${\cal M}_2$ and 0.316 for ${\cal M}_3$. This shows a fact which is that often
the unconstrained model will perform better on dynamical data but far worse on static
data. Hence, when dynamical and static performances are conflicting objectives (more on
this later), it is possible to give up a bit of dynamical performance to improve the
model steady-state, as for ${\cal M}_3$ in this example.

\subsubsection{Biobjective parameter estimation}

In the previous section the auxiliary information \I was coded in terms of equality constraints.
Hence such hard constraints were the means by which \I found its way into the model. In this
section a different approach will be described.

For the sake of presentation, only the biobjective case will be described, although more objectives
can be taken into account \citep{nep_eal/07}.

In biobjective optimization we do not search for a {\it single optimal}\, solution
that simultaneously minimizes two cost functions. Instead the aim is to find
a set $\Theta^*$ of solutions $\btheta^*$ called the Pareto set  \citep{cha_hai/83}
\begin{equation}
\label{pareto} 
\btheta^* \in \Theta^* \Leftrightarrow \{\exists\!\!\!/~ \btheta
: ~{\bf J}(\btheta) \le {\bf J}(\btheta^*)~{\rm and}~ {\bf J}(\btheta)
\neq {\bf J}(\btheta^*) \},
\end{equation}
\noindent
where ${\mathbf J}(\btheta)$ is a vector of cost functions. In (\ref{pareto})
the symbol ``$\le$'' indicates that all elements of a vector are less or equal
to the corresponding elements of the other vector. An often-used choice is:
\begin{eqnarray}
\label{eq:vector}
{\mathbf J}(\btheta) = \left[ \begin{array}{cc} J(Z,\,Z_{{\cal M}_1}) &
J(Z_{\rm ss},\,Z_{{\cal M}_{\rm ss}})
\end{array} \right]^{\T} ,
\end{eqnarray}

\noindent
where $Z_{{\cal M}_{\rm ss}}$ is the data produced by model ${\cal M}$ taken
in steady-state. The solutions $\btheta^*$ can be found solving \citep{nep_eal/04,nep_eal/07}
\begin{equation}
\label{paretoset}
\begin{array}{lcl}
\hat{\btheta}_{\rm BO} = & {\rm arg}~~{\rm min} &
\lambda J(Z,\,Z_{{\cal M}_1}) + (1-\lambda)J(Z_{\rm ss},\,Z_{{\cal M}_{\rm ss}}) , \\
~ & \btheta \in \mathbf{D} & ~\\
\end{array} 
\end{equation}

\noindent
where $0 \leq \lambda \leq 1$ and $\mathbf{D}$ is the set of viable solutions.
For each value of $\lambda$ one estimates $\btheta^*(\lambda)$, which is part of
the Pareto set. The mono-objective solutions that minimize $J(Z,\,Z_{{\cal M}_1})$ and
$J(Z_{\rm ss},\,Z_{{\cal M}_{\rm ss}})$ are obtained from (\ref{paretoset}) 
$\lambda = 0$ and $\lambda = 1$, respectively.  An interesting interpretation of  (\ref{paretoset})
is that it is the Least Squares estimator with a regularization term taken from the
static data, which is the available auxiliary information in this case.

In order to provide some details on how this can be achieved, the following results are
required:

\begin{definition}\emph{Affine Information}  \citep{nep_eal/07}.
Consider the parameter
  vector~$\btheta \in \bbR^{n_\theta}$, a vector $\mathbf{v} \in \bbR^p$ and a
  matrix $G \in \bbR^{p \times {n_\theta}}$. Both $\mathbf{v}$ and $G$ are assumed
  to be accessible.  Moreover, suppose $G \btheta$ constitutes an
  \emph{estimate} of $\mathbf{v}$, such that $\mathbf{v} = G \btheta +
  \epsilon$, where $\epsilon \in \bbR^p$ is an error vector.  Then
  $[\mathbf{v},G]$ is said to be an  \emph{affine information}
  pair of the system.
\end{definition}

\begin{theorem}\citep{nep_eal/07}
Let $[{\bf v}_i,G_i]$ with $i = 1, \ldots, m$ be $m$ affine information 
pairs related to a
system, where $\mathbf{v}_i \in \bbR^{p_i}$ and $G_i \in
\bbR^{p_i \times {n_\theta}}$. Assume that at least one of the
matrices $G_i$ is full column rank.
Let $\mathcal{M}$ be a given model structure which is linear
in the parameter vector $\btheta \in \bbR^{n_\theta}$. Then the $m$ affine
information pairs can be simultaneously taken into account while estimating
the parameters of model $\mathcal{M}$, by solving
\begin{equation}
\label{minj}
\hat{\btheta}_{\rm MO} = {\rm arg~} \mathop{\rm min~}_{{\hat{\theta}}}
\sum_{i=1}^{m}w_i (\mathbf{v}_i - G_i\btheta)^{\T}(\mathbf{v}_i -
G_i\btheta),
\end{equation}
with ${\bf w} = [ \begin{array}{ccc} w_1 & \ldots & w_m \end{array} ]^{\T} 
\in W$.  The unique solution of (\ref{minj}) is given by
\begin{equation}
\hat{\btheta}_{\rm MO}  =  \left[\sum_{i=1}^{m}w_iG_i^{\T}
G_i\right]^{-1}\left[\sum_{i=1}^{m}w_iG_i^{\T} \mathbf{v}_i
\right].
\end{equation}
\end{theorem}

In (\ref{paretoset}) the multiobjective problem has been reduced to a biobjective one ($m=2$)
with $w_1=\lambda$ and $w_2=1-\lambda$, $G_1=\Psi$ is the matrix of ``dynamical''
regressors and $G_2$ would be composed of regressors taken in steady-state. Also,
$\mathbf{v}_1=\by$ and $\mathbf{v}_2=\bar{\by}$.

Once the Pareto set is obtained, varying $\lambda$ in the range $0 \leq \lambda \leq 1$,
it is often desirable to pick one model, one way of doing this was proposed by
\cite{bar_eal/07}. Each candidate model $\btheta$ in the Pareto set  
$\Theta^*$ is simulated in free-run mode and the output is indicates as: 
\begin{equation}
\label{eqn:model_simulation}
\hat{y}(k) = \hat{\bpsi}^{\T}(k-1)\btheta,
\end{equation}
where the hat in $\hat{\bpsi}(k-1)$ indicates that previously simulated values are used.
The corresponding error is:
\begin{equation}
\label{eqn:error}
\eta(k) = y(k) - \hat{y}(k).
\end{equation}
Now, define
\begin{equation}
\label{eq:correlation_nova} 
J_{\rm corr}(\btheta) = \left| \frac{1}{N}\sum_{k=1}^N \eta(k) \hat{y}(k) \right|.
\end{equation}
Barroso has argued that $J_{\rm corr}(\btheta)$ will achieve its lowest value
for the best $\btheta$ \citep{bar_eal/07}. 

An interesting remark concerning multiobjective estimation in grey-box system identification
is that it provides a mechanism by which auxiliary information \I can be taken into account.
Not in the form of constraints, but rather by defining a cost function for which smaller values
correspond to models that better incorporate \I$\!\!$. Since one of the objectives will still be 
to fit the dynamical data, then various types of auxiliary information can be used to compose
a multiobjective optimization problem. In this section bi-objective optimization was used
in the context of parameter estimation. Hafiz has investigated approaches
in the context of both, structure selection and parameter estimation \citep{haf_eal/19,haf_eal/20}.

\subsection{Other types of auxiliary information}

In what follows other  types of auxiliary information \I that were used in the context of grey-box
identification will be listed.

One of the basic forms of \I is that of  the shape of the static function. The use of such
information has been used in \citep{agu/97pd,agu_eal/99i3e}. Another basic form of
\I is the location of fixed-points. As a matter of fact, as discussed in this work, the static
curve of a system can be interpreted as a {\it continuum}\, of fixed points as a function of
the input. Relationships of how the number, location and symmetry of fixed points relate to
the structure of polynomial models have been presented in \citep{agu_men/96}.

A key point in this way of using \I is to establish relationships between model structure and
types of information. Some results for polynomial NARX models were developed in
\citep{agu_jac/98,agu_eal/02iee}. Static functions were used for grey-box identification of
rational models \citep{cor_eal/02}, polynomial models \citep{agu_eal/04iee} and 
radial basis models in \citep{agu_eal/07}.

Symmetry of the flow in state space was a key piece of information used to solve certain
modelling problems via grey-box identification \citep{agu_eal/04,agu_eal/08solar} and in the design of 
systems \citep{che_eal/08}.

Because the normal forms of bifurcations in maps are polynomial, certain
aspects of bifurcations of equilibria can be used in \I$\!\!$. In particular, the {\it flip}, or period-doubling
bifurcation was used in \citep{agu_fur/07}, Hopf (Neimark-Sacker) bifurcation was imposed
on identified maps in \citep{agu_rod/12} and the transcritical bifurcation was used to reproduce
 non-smooth static functions in \citep{agu/14}.
 
 For systems that are known to display hysteresis, a specific choice of inputs can be used
 to guarantee multi-stability \citep{mar_agu/16}, which is a condition for hysteresis.  More
 recently, an additional constraint was shown to be fundamental in order to guarantee that the
 system remains on the hysteresis loop in the case the input becomes constant \citep{abr_eal/20}.
 In the first
 case study discussed in Section~\ref{eps}, a step further will be given: not only will the
 appearance of the hysteresis loop be guaranteed by the use of a specific class of inputs, but
 also, the shape of the loop will be modified by the use of parameter constraints.
 
 A somewhat different approach has been followed in \citep{wu_eal/20}. In that work the
 authors use two sets of data of the same type, e.g. both coming from dynamical tests.
 However, the knowledge about the system conveyed by each data set is different. Then, 
 one of the data sets is considered \I to the next. 

The problem of using equality constraints on the states has been considered in \citep{ric_tei/21}.
In that paper, the problem has been solved in two steps: first the constraints are mapped from 
state variables to matrix parameters and then the problem is vectorized in order to apply standard algorithms. 
The region of attraction of the equilibrium point has been used as auxiliary information -- called side information -- in
\citep{kho_smi/21}.

\section{Examples and Case Studies}
\label{eps}

The literature that has been cited so far provides many examples -- simulated and
experimental -- of identification problems. Here three examples will be briefly presented
for the sake of illustration with some emphasis on grey-box solutions.

\subsection{Pneumatic valve with hysteresis}

The data used in this example have been collected by Petrus Abreu, Lucas Tavares and
Guilherme Mello from a laboratory pilot plant. Previous studies of the dynamics and steady-state
features have been reported elsewhere \citep{agu/14,rib_agu/18}. 
Valves are widely used in industrial process control although they suffer from
loss of performance due to nonlinearities and friction. For this reason the 
compensation of such effects has been investigated \citep{rom_gar/11,bae_gar/18,abr_eal/20,tav_eal/21}.

The aim in this example is to illustrate results from \citep{mar_agu/16} and Sec.\,\ref{subssd} 
and to show how a hysteresis loop may be obtained using NARX polynomial models. 
In the sequel, it will be shown how the use of
constraints can be employed to alter the format of the obtained loop.

\begin{figure}[t!]
\begin{center}
\includegraphics[scale=0.4]{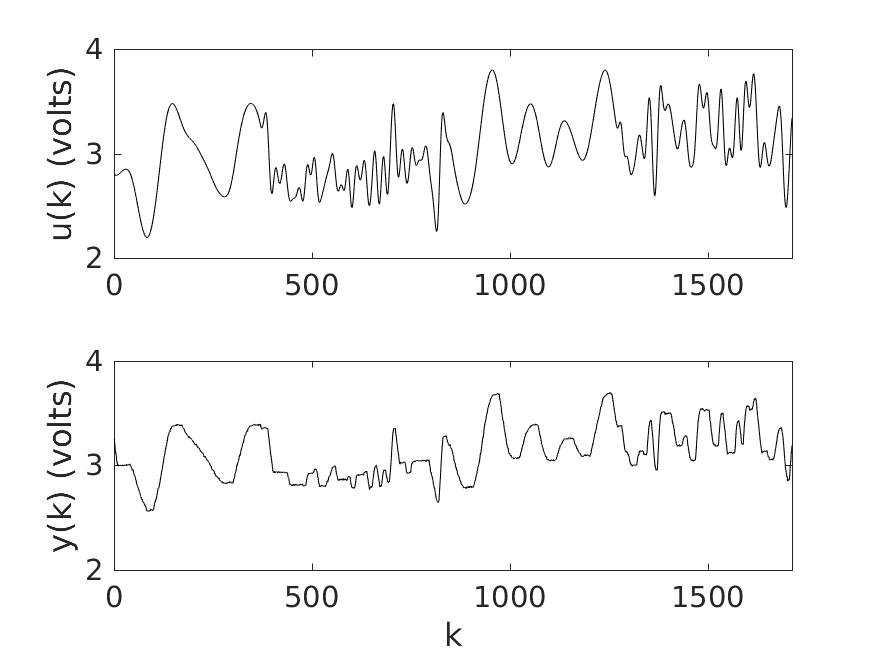}  
\caption{\small Input (top) output (bottom) data from a pneumatic valve. The
input is a voltage signal sent to a V/I converter, and subsequently to an I/P
converter which manipulates a pressure source. The output, also a voltage, is the movement of
the actuator stem, collected by a position sensor.}  
\label{valvedata}                               
\end{center}                                
\end{figure}

The identification data are shown in Figure~\ref{valvedata}. The input and output 
are voltage signals, as indicated in the figure and explained in the caption. The units
of such signals will be omitted henceforth. Such data were collected at a sampling frequency of 100\,Hz and
later decimated by a factor of 10. The identification data length was $N=1700$.

\begin{figure}[b!]
\begin{center}
\includegraphics[scale=0.4]{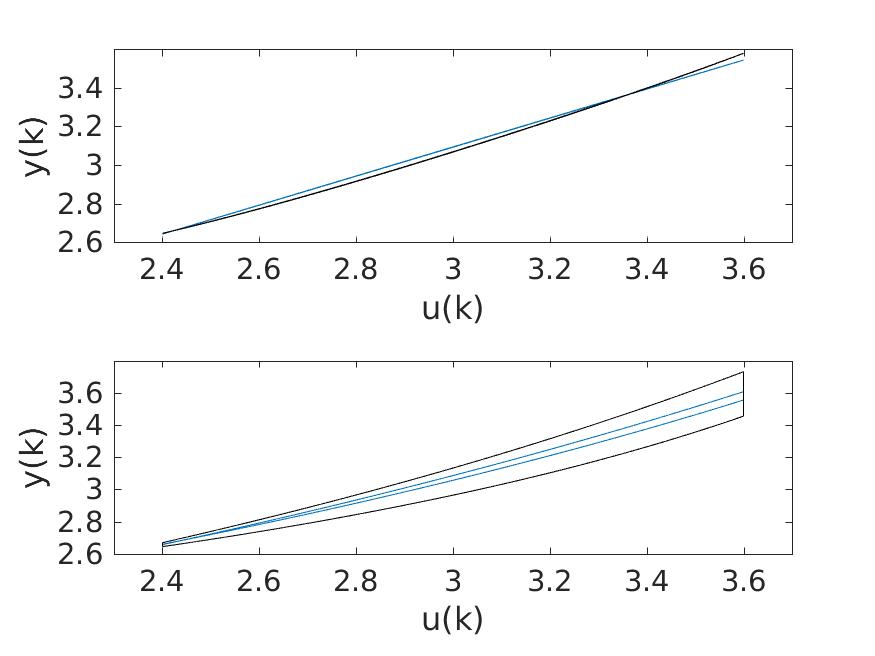}  
\caption{\small Top: static functions for three-term affine model (straight line in light blue) and thirteen-term
nonlinear black-box model (slight curve in black). Bottom: static functions with hysteresis loops for
model (\ref{a130519}) (thin loop in light blue) and for a parameter-constrained model (wider
hysteresis loop in black).}  
\label{hysteresis}                               
\end{center}                                
\end{figure}

Four models will be mentioned in what follows. We start with a very simple model
which has three regressors: $y(k-1)$, $u(k-1)$ and the constant. Such a model
presents a mean absolute percentage error of about MAPE~$\approx 6\%$, but being an affine model
it is not expected to reproduce anything that resembles hysteresis. In fact, following
the discussion by Prof. Bernstein, this can be checked by exciting the model with a
loading-unloading signal of decreasing frequency. As the input frequency diminishes
any loops seen on the $u(k)-y(k)$ plane will collapse to a single curve \citep{ber/07}.
This is seen in the top of Figure~\ref{hysteresis} where the static function of this
three-term model is just a line, as expected.

The second model considered is a thirteen-term nonlinear model with up to cubic
nonlinear terms ($\ell=3$) and maximum lags of two $n_u=n_y=2$. Although this model
presents a slightly better dynamical performance with  MAPE~$\approx 4.7\%$, it
does not reproduce any hysteresis behavior. This is seen by noticing that as the frequency of the
input forcing reduces, the model settles to a static function which is a slight curve
as seen in the top of Figure~\ref{hysteresis}. It has been pointed out that more often than
not, models obtained in a black-box fashion do not present hysteresis \citep{mar_agu/16}.
In what follows, certain modeling decisions -- which clearly constitute grey-box procedures --
will be made in order to  guarantee that the identified models will have hysteresis.

The third model considered in this example is shown below:
\begin{eqnarray}          
\label{a130519}      
 y(k) & = & +0.80665 \, y(k-1) +0.02888 \, u(k-1) +0.30362 \nonumber \\
     & &   +0.57737 \, u_2(k-1)u(k-1) -0.52294 \, u_2(k-1)y(k-1)\nonumber \\          
      & &   +0.022105 \, u(k-1)y(k-1) -0.00864 \, u_3(k-1)u(k-1)\nonumber  \\
      & &  +0.00787 \, u_3(k-1)y(k-1)  ,
\end{eqnarray}     

\noindent
where $u_2(k)$ is the first difference of the input signal, that is, $u_2(k)=u(k)-u(k-1)$
and $u_3(k)={\rm sign}[u_2(k)]$. The static behavior of model (\ref{a130519}) is 
attained by assuming a constant input $\bar{u}=u(k),~\forall k$, and $u_2(k)=0$.
In addition to that, for loading periods $u_3(k)=1$ and for unloading $u_3(k)=-1$.
The use of regressors in $u_3(k)$ is recommended for hysteretic systems \citep{mar_agu/16}.
Hence the decision of including $u_2$ and especially $u_3$ was based on the 
auxiliary information \I that the system presents hysteresis. This, of course, constitutes a
grey-box procedure. Although the dynamical behavior has not changed significantly
(MAPE~$\approx 5\%$), the novelty is that model (\ref{a130519}) does have a hysteresis loop,
as shown in the bottom part of Figure~\ref{hysteresis}. In fact such a model has
two equilibria, one for loading, given by
\begin{eqnarray}          
\label{b130519}      
 \bar{y} = \frac{(\theta_2+\theta_7)\bar{u}+\theta_3}{1-\theta_1-\theta_8-\theta_6\bar{u}} 
\end{eqnarray}    

\noindent
and one for unloading:
\begin{eqnarray}          
\label{c130519}      
 \bar{y} = \frac{(\theta_2-\theta_7)\bar{u}+\theta_3}{1-\theta_1+\theta_8-\theta_6\bar{u}} ,
\end{eqnarray} 

\noindent
where the $\theta$'s are given in the same order as that of the parameters in model  (\ref{a130519}).

\begin{figure}[b!]
\begin{center}
\includegraphics[scale=0.3]{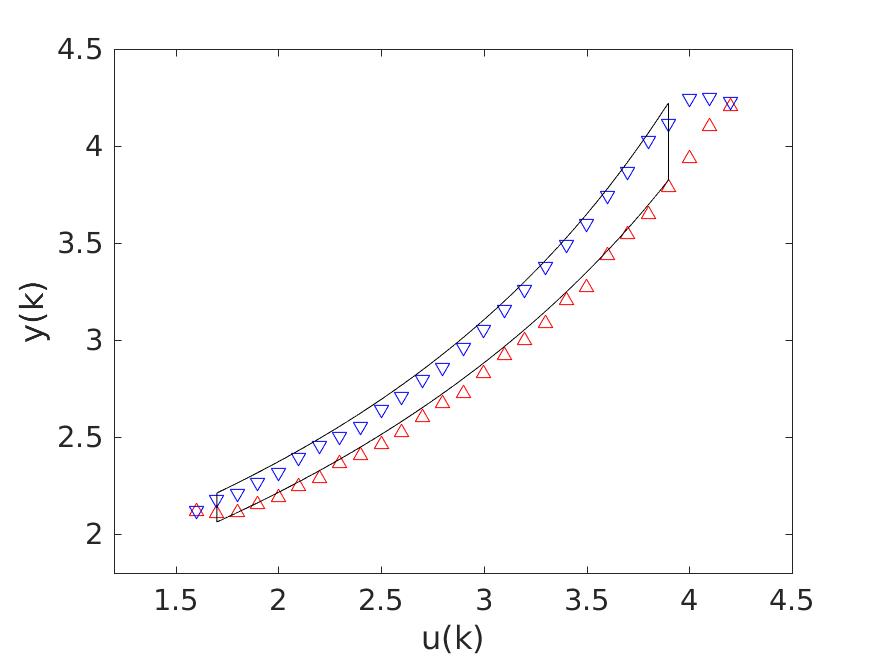}  
\caption{\small Static function with hysteresis loops for the constrained
model (continuous lines) and valve data collected in steady-state for
 \textcolor{red}{(red) $\bigtriangleup$} increasing inputs, and \textcolor{blue}{(blue) $\bigtriangledown$} decreasing inputs.}  
\label{hysteresis2}                               
\end{center}                                
\end{figure}

Now because we have valve data collected in steady-state, this is the auxiliary information \I in this
problem, we would like to make sure that the identified model has a static function with a hysteresis
loop that better resembles the static data. This can be achieved by building parameter constraints, as discussed in
Sec.\,\ref{sscp}, from such data. In particular we write
\begin{equation}
{\bf c}=
\left[\begin{array}{c}
     \bar{y}^1\\
     \bar{y}^2 \\
     \bar{y}^3\\
     \bar{y}^4 
 \end{array}
\right] ;
\hspace{0.5cm}
S=
\left[\begin{array}{cccccccc}
\bar{y}^1& \bar{u}^1& 1 & 0 & 0 &\bar{u}^1\bar{y}^1 & \bar{u}^1& \bar{y}^1\\
\bar{y}^2& \bar{u}^2& 1 & 0 & 0 &\bar{u}^2\bar{y}^2  & \bar{u}^2& \bar{y}^2 \\
\bar{y}^3& \bar{u}^3& 1 & 0 & 0 &\bar{u}^3\bar{y}^3 & -\bar{u}^3& -\bar{y}^3\\
\bar{y}^4& \bar{u}^4& 1 & 0 & 0 &\bar{u}^4\bar{y}^4  & -\bar{u}^4& -\bar{y}^4
\end{array}
\right], \nonumber
\end{equation}

\noindent
where the superscripts indicate a specific data point, that is, 
$(\bar{u}^1,\,\bar{y}^1)=(1.8,\,2.112)$, $(\bar{u}^2,\,\bar{y}^2)=(3.4,\,3.249)$,
$(\bar{u}^3,\,\bar{y}^3)=(1.7,\,2.211)$ and $(\bar{u}^4,\,\bar{y}^4)=(2.7,\,2.843)$
are four desired points, the two first on the loading part of the hysteresis loop, and the
other two on the unloading side. Then, using estimator (\ref{constrained}) a new set of 
parameter values is attained and the hysteresis loop is wider, as seen in 
Figure~\ref{hysteresis2} and has (MAPE~$\approx 6\%$). The slight increase in the value
of MAPE is somewhat expected because as we have imposed four steady-state constraints
there are less degrees of freedom to fit the dynamical data.

As seen in the bottom part of Figure~\ref{hysteresis}, the hysteresis loop of the constrained
model is indeed wider than when the auxiliary information was not used. Besides, the static
function of the constrained model compares favorably to the collected data, as seen in 
Figure~\ref{hysteresis2}.

It has been verified that for some model-based control strategies applied to systems
with hysteresis the use of auxiliary information was important. In particular, the
control laws implemented from grey-box models clearly outperformed those based on
black-box models \citep{abr_eal/20}.

The aim of this example has been to illustrate the use of grey-box techniques in
dealing with data measured from a system known to have hysteresis. The next two
examples illustrate other aspects of grey-box techniques in other applications.

\subsection{Dynamical and steady state data from an oil well}

This case study has been developed in cooperation with colleagues in academia and industry \citep{agu_eal/17soft}.

``Downhole pressure is an important process variable in the operation of gas-lifted oil wells. 
The device installed in order to measure this variable is often called a Permanent Downhole 
Gauge (PDG). Replacing a faulty PDG is often not economically viable and to have
an alternative estimate of the downhole pressure is an important goal'' \citep{agu_eal/17soft}.
One way around this problem is to develop softsensors, that is, build models that
will estimate the variable of interest (downhole pressure) from readily available measurements.
Of course, this means that the models must be identified before the PDG becomes faulty.
After the PDG becomes inoperative, a valid model can be used to provide an estimate
for the said pressure.
More details about this application can be found in \citep{tei_eal/14,agu_eal/17soft}.

This challenge was faced exclusively from historical data, that is, no specific testing was
performed. If, on the one hand, this sounds an advantage, on the other, there is a cost to
pay for that. In historical data most of the time the process is in steady state. Hence, in 
order to find windows of data that will bring forth the process dynamics, special care must
be taken. The inspection of the database by an expert in the search for useful transients
is both slow and tedious. In order to overcome this, an automatic procedure was developed
and used \citep{rib_agu/15}. The authors of \citep{pit_eal/20} have investigated the impact of
the window of data on the performance of the identified model. They have put forward two
procedures to select adequate windows of data in the context of linear thermal models. 

Not everything is bad in historical data of this sort. There is a lot of steady state information.
In fact, it is possible to attain high quality steady-state data by averaging over rather long
windows of data. As discussed in Section~\ref{gb} steady-state information can be used
to great advantage in the identification of {\it dynamical}\, models, so long it is treated as
auxiliary information \I which is additional to a set of dynamical data $Z$.

\begin{figure}[b!]
\begin{center}
\includegraphics[scale=0.3]{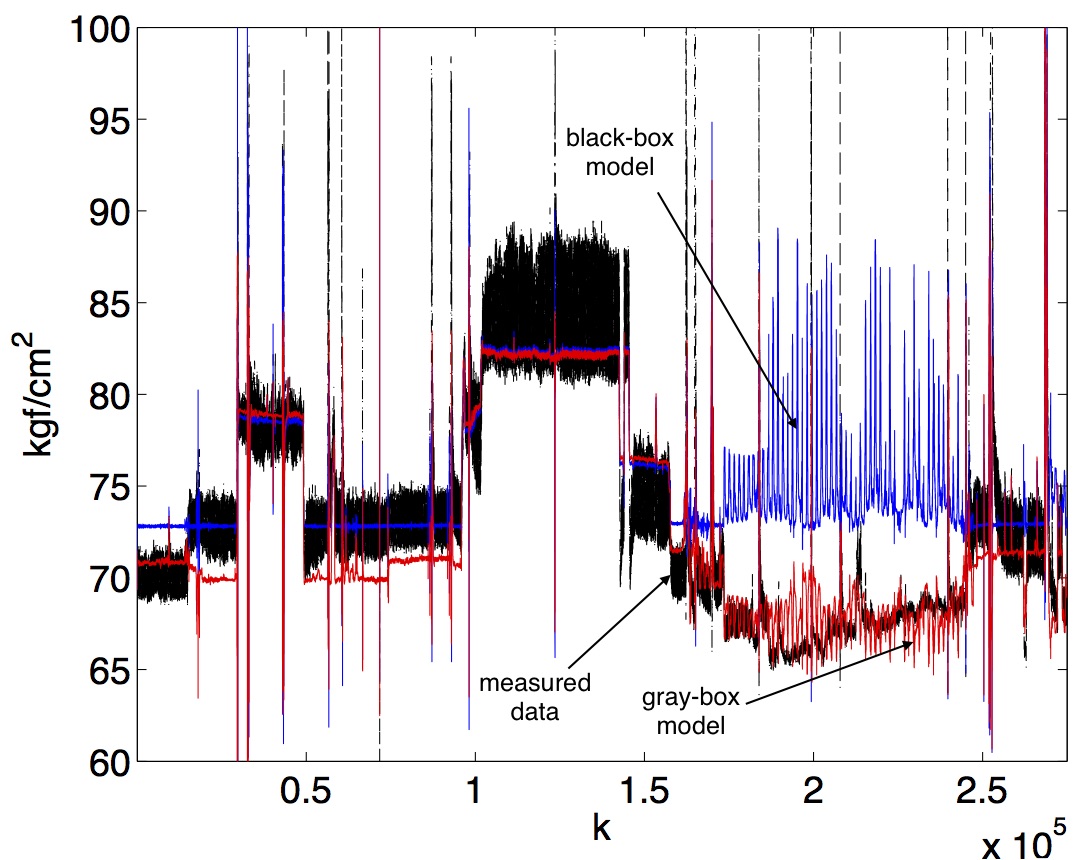}  
\caption{\small Black is the measured downhole pressure, blue is the output of a black-box polynomial model
and red is the response of a grey-box polynomial model. See \citep{agu_eal/17soft} for details.}  
\label{black_grey}                               
\end{center}                                
\end{figure}

As an example of a SISO (single-input single-output) polynomial model, a six-term model was estimated using
the following metaparameters:
$\ell=2$, $n_y=n_u=3$ and $n_e=1$, with downhole pressure as output and gas-lift pressure 
as input. A transient window $Z^N$ of length $N=2000$ (around $k\approx 1.0\times 10^5$ in Figure~\ref{black_grey}, the sampling time
was $T_{\rm s}=1$\,min) was used, which corresponds to 33.3 hours of operation. 
The six regressors that compose the model were chosen from a set of 220 candidate
terms using the ERR \citep{bil_eal/89} (see Sec.\,\ref{errsrr}) and term clustering \citep{agu_bil/94b}.
Parameters were estimated using orthogonal extended least squares. 
This resulted in a set of models with similar performance on validation data. Based on the principle of
parsimony,  the simpler model was retained \citep{agu_eal/17soft}.

\begin{figure}[hb!]
\begin{center}
\hspace{-0.3cm} \includegraphics[scale=0.25]{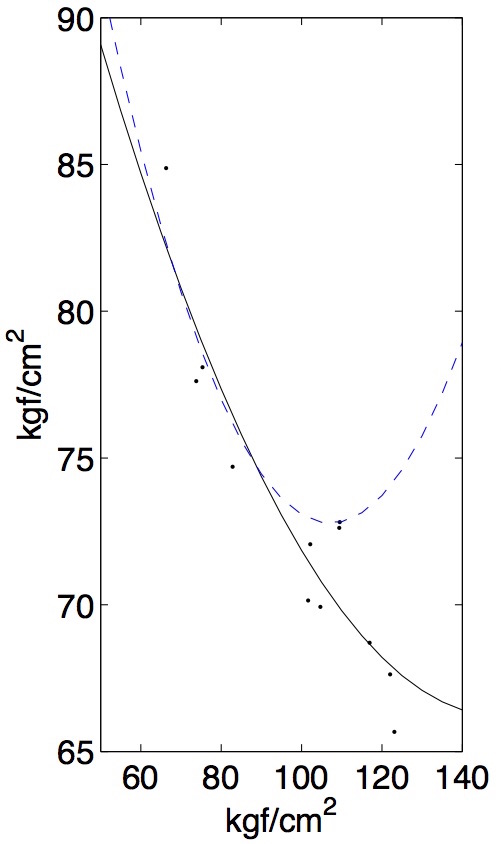}  
\caption{\small The dots indicate steady state data: $x$-axis is gas-lift pressure and $y$-axis is
downhole pressure. The (black) solid line is a second order polynomial estimated
from the data which ideally should be closely followed by that of estimated models.
The blue dashed curve corresponds to the static characteristic of a black-box polynomial model
which is only accurate in the range of the dynamical data (around 80\,kgf/cm$^2$).
See \citep{agu_eal/17soft} for details.}  
\label{fig:ssdata}                               
\end{center}                                
\end{figure}

It is important to notice that the identification data being around $k\approx 1.0\times 10^5$ in Figure~\ref{black_grey}
means that they were measured when the process was operating roughly in the range of downhole pressure
75--80\,kgf/cm$^2$. Over that  range, the static nonlinearity of the model is quite consistent 
with the steady-state data (see Figure~\ref{fig:ssdata}). Right after the identification window the process
operate over a month (34 days) around 85\,kgf/cm$^2$ slugging from 80 to 90\,kgf/cm$^2$. After this,
operation reduced the pressure gradually to levels as low as 70\,kgf/cm$^2$. Around this point
the process static behavior is quite different from that in the identification window. Hence it is no
surprise that the black-box model is unable to correctly reproduce such a behavior at low pressure
values.

One way around this problem is by means of grey-box modeling, as discussed in Section~\ref{gb}.
Hence steady-state data over a wide range of operating points were taken fro?m the historical 
records by computing the time average over long windows of data whenever the process operated at
a single point. Some of the averaged steady-state data are shown in Figure~\ref{fig:ssdata}.
Now using such data {\it simultaneously}\, with the original dynamical data, grey-box models were
estimated by means of constrained estimation (see Section~\ref{gb}). As a consequence, the resulting
models have the dynamics underlying the original data set $Z^N$ which is roughly in the range 
75--80\,kgf/cm$^2$ and the static behavior of the data in Figure~\ref{fig:ssdata}. The performance
of one of such models against the black-box model is shown in Figure~\ref{black_grey}, which 
corresponds to over six months of operation. The
benefits of using steady-state data in this nontrivial example is evident.

\begin{figure}
\centering
\begin{tabular}{c}
(a)  \\ 
\includegraphics[scale=0.4]{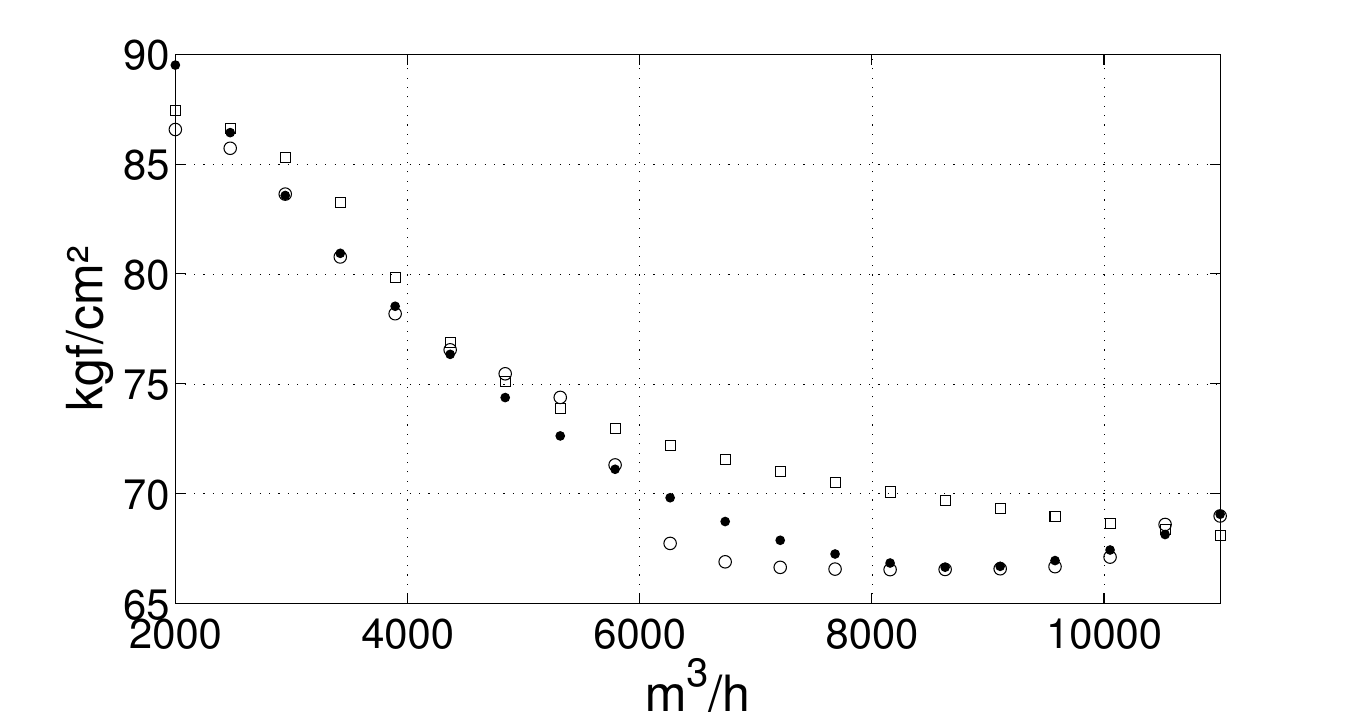}\\ (b) \\
\includegraphics[scale=0.4]{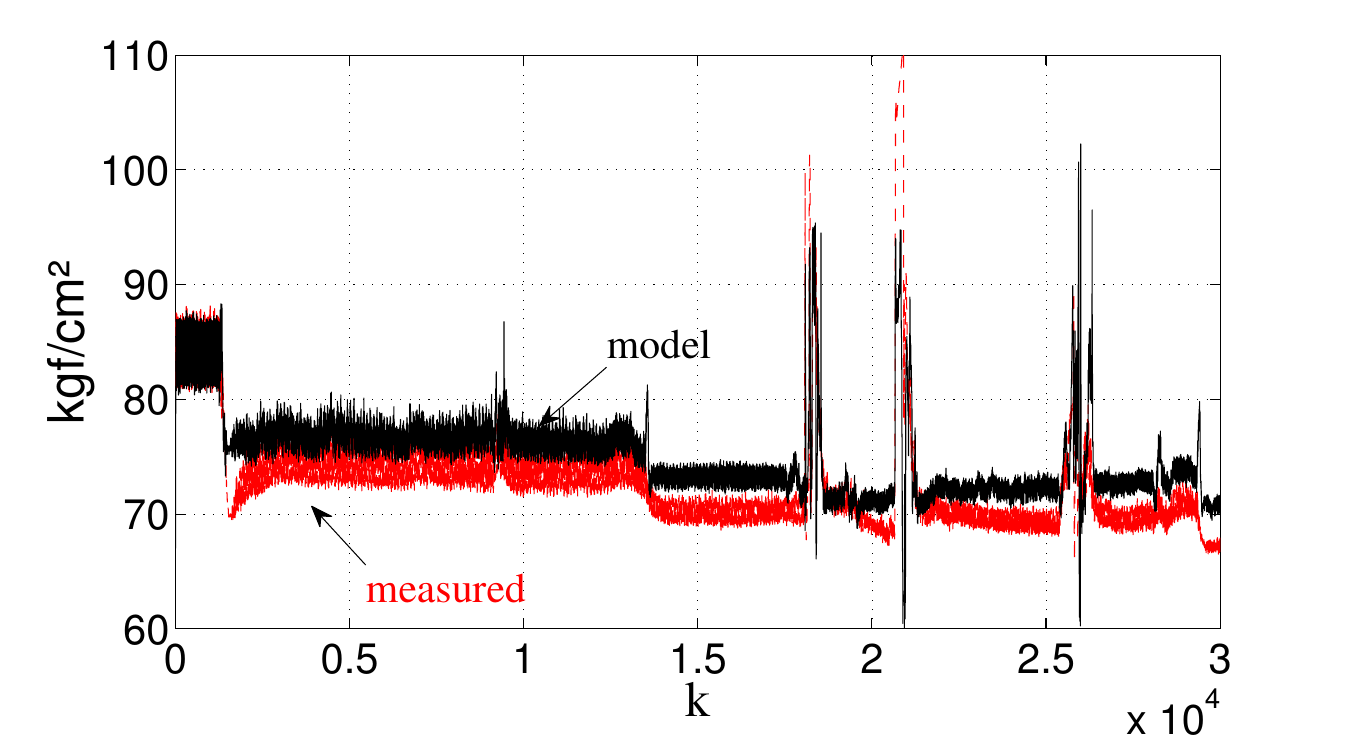}  \\ (c)\\
\includegraphics[scale=0.4]{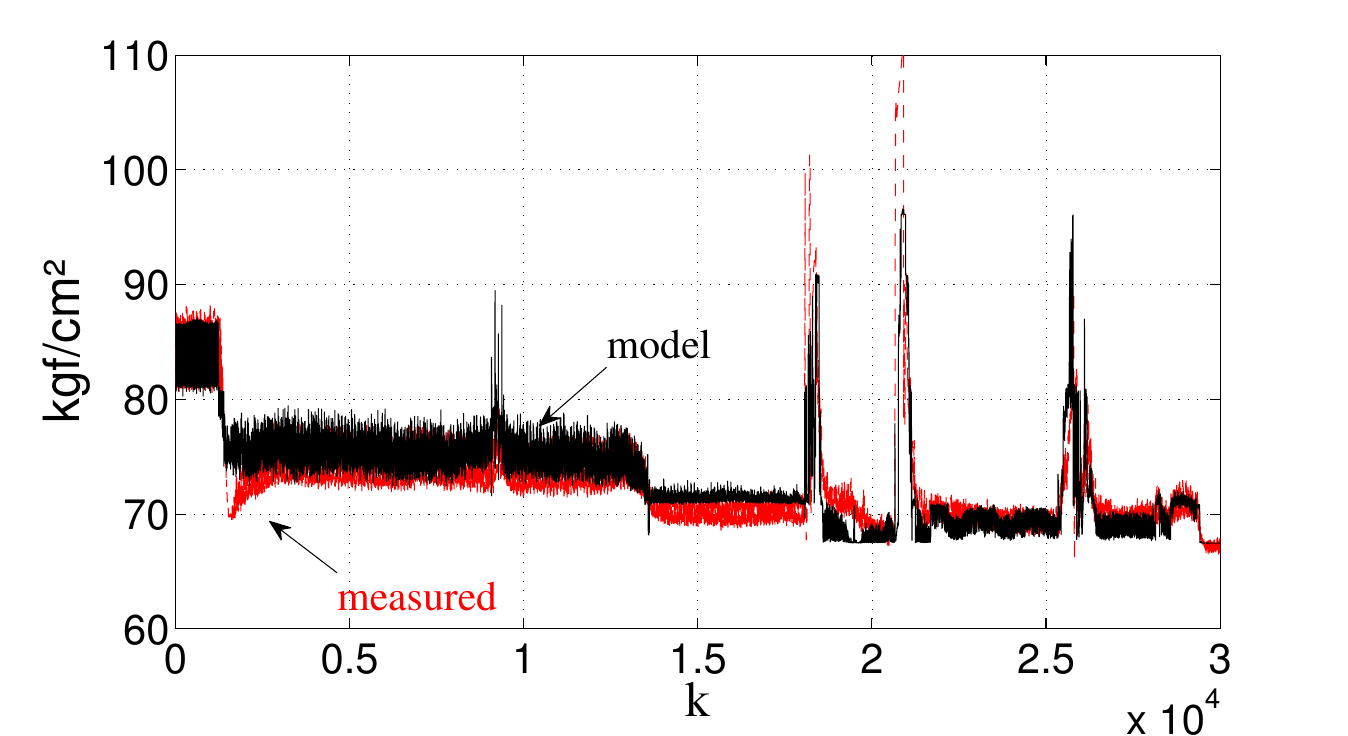}    
\end{tabular} 
\caption{\small \label{fig:static}(a)~Static curves (gas-lift flow \textit{versus} downhole pressure), ($\bullet$) desired, ($\square$) black-box model and ($\circ$) grey-box model;  downhole pressure prediction over validation data using Neural Networks: 
(b) black-box model and (c) grey-box-model~\citep{fre/13}. }
\end{figure}

Similar results have been reported for neural networks \citep{agu_eal/17soft} as shown in
Figure~\ref{fig:static}.

\subsection{Imposing equilibria to model vector fields}

This case study has been developed in cooperation with Rafael dos Santos \citep{san/18tese} and Guilherme Pereira \citep{san_eal/18}.

\subsubsection{The challenge}

\begin{figure}[htb]
\centering
\includegraphics[scale=0.6]{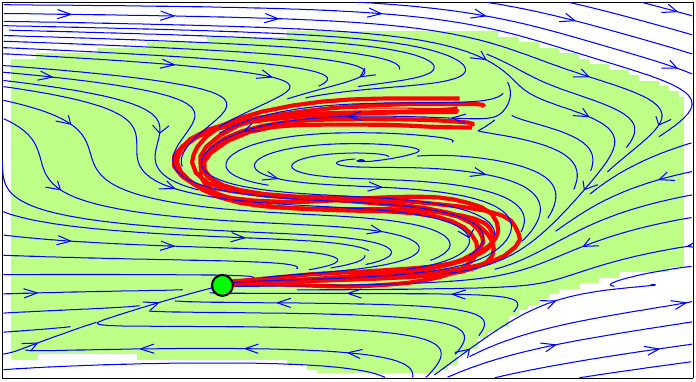}
\caption{\small \label{fig1}The vector field is represented by the blue lines with arrows. A set of possible trajectories are shown in red.
The green region is the basin of attraction of the target, indicated by the green circle \citep{san_eal/18}. }
\end{figure}

An interesting problem in robotics is that of programming a robot to reach a certain location or position. A basic
way of achieving this is to provide a trajectory to be followed. How to define which trajectory to provide for the
robot is another issue.  An elegant alternative is not to provide a trajectory itself but rather to provide a vector field
for which infinite trajectories can be obtained, one for each possible initial condition. Hence, a vector field
can be seen as a trajectory-producing mechanism. However for the resulting ``solution'' to be a useful trajectory
it must end at the target. Traducing this in nonlinear dynamics language, the target must be a stable fixed point
and the initial conditions must be taken within the corresponding basis of attraction. This is illustrated in 
Figure~\ref{fig1} where any initial condition taken in the green area will result in a trajectory that will eventually
converge to the attractor, shown as a green circle. A set  of trajectories that start at a set of neighbouring initial conditions are shown.
Initial conditions taken in the white region will diverge to some other part of the state space.

\begin{figure}[htb]
\centering
\begin{tabular}{cc}
(a) & (b) \\
\includegraphics[width=0.35\textwidth]{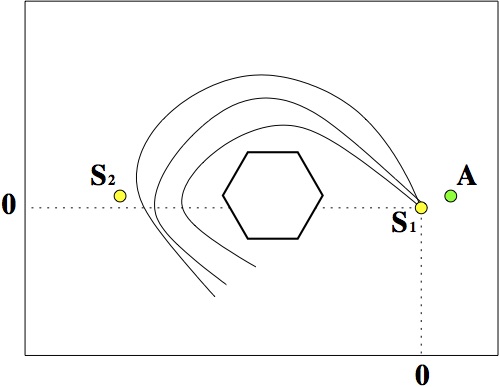} &
\includegraphics[width=0.35\textwidth]{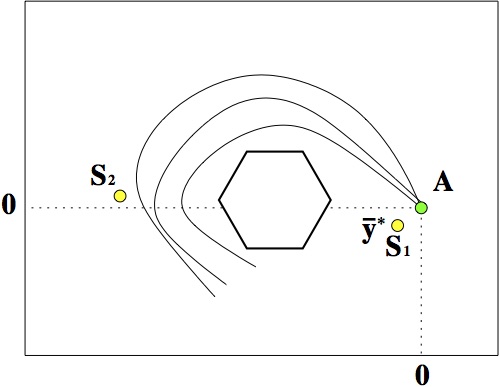}
\end{tabular}
\caption{\label{fig:modelo}\small Three trajectories provided by the teacher. All the trajectories should finish at the target A that must
be a stable fixed point of the resulting vector field (see text for details). In providing the trajectories, the teacher can, 
for instance, avoid obstacles, indicated here by the polygon. $S_2$ is a saddle that is not used in the example. Typical scenarios for 
(a)~black-box techniques, and (b)~grey-box methods. In (b) the {\it location} of $S_1$ was imposed, that is $S_1=\bar{\by}^*=[y_1^*, y_2^*]^{\T}$ is the new position of $S_1$, and the origin remains a fixed point. The stability type of $S_1$ and $A$ did not change. See text and  \citep{san_eal/18} for details.}
\end{figure}

The goal in this case study is to identify from data a model for the vector field. Hence, such a model
can be used to produce trajectories for the robot to follow. The first question we should ask is where do the data
come from. In this class of problems the data come from demonstrations, that is, a ``teacher'' (usually a
human) performs a set of valid trajectories which are recorded and used as identification data (Fig.~\ref{fig:modelo}).

A central point in this problem can be stated as a question:  is it guaranteed that the identified
model will have a single attractor and that it will appear exactly at the meeting point of the
trajectories? This question has a double answer. If the identification follows a black-box
procedure, then the answer is negative (Fig.~\ref{fig:modelo}a). However, using grey-box techniques it is possible
to guarantee that the model will satisfy the design requirements (Fig.~\ref{fig:modelo}b). This will be briefly detailed
in what follows. Further details can be found in \citep{san_eal/18}.

\subsubsection{The methodology}

For the sake of simplicity, the 2D will be described. The formulation for higher-order systems
is straightforward although the implementation could become infeasible for such systems \citep{san_eal/18}.

Given a set of $n$ demonstrations $\mathbf{Y}_{(n)}$,  which are trajectories in a state space $\bbR^2$, the
aim is to estimate NAR models that will approximate the vector field of which $\mathbf{Y}_{(n)}$ are
integral curves. In this case, $\by(k)=[y_1(k) \ y_2(k)]^{\T}$, where
$y_1(k)$ and  $y_2(k)$ are the coordinates in $\bbR^2$.
Hence we search for models composed of 2 first-order difference equations, to produce $\hat{y}_1(k), \hat{y}_2(k)$. 

The pool of candidate regressors is composed by all possible combinations up to  degree $\ell$ of such variables plus a constant term.
The regressors of each model equation are automatically chosen using the ERR criterion (see Eq.\,\ref{eq_errp}) 
together with Akaike's criterion \citep{aka/74}. A typical model has the general form:
\begin{eqnarray}
\label{a070116}
y_1(k)& =& F_1^\ell[y_1(k\!-\!1),\ y_2(k\!-\!1)]\!+\!e_1(k) \nonumber \\
y_2(k)& =& F_2^\ell[y_1(k\!-\!1), \ y_2(k\!-\!1)]\!+\!e_2(k) 
\end{eqnarray}

Assuming the model is stable, in steady-state $y_1(k)=y_1(k-1)=\bar{y}_1$; $y_2(k)=y_2(k-1)=\bar{y}_2$ and
dropping the noise, the equilibria of 
model (\ref{a070116}) -- called fixed points -- are given by $(\bar{y}_1,\,\bar{y}_2)$ that are the solutions to the set of equations:
\begin{eqnarray}
\label{b070116}
\bar{y}_1& =& F_1^\ell[\bar{y}_1,\,\bar{y}_2] \nonumber \\
\bar{y}_2& =& F_2^\ell[\bar{y}_1,\,\bar{y}_2] 
\end{eqnarray}

\noindent
and the stability of such fixed points can be determined by the eigenvalues $\lambda_1, \lambda_2$ of the Jacobian matrix 
\begin{eqnarray}
\label{a130116}
{\rm D} F (\by)= \left[
\begin{array}{ccc}
\displaystyle \frac{\partial F_1^\ell}{\partial y_1(k-1)} &  \displaystyle \frac{\partial F_1^\ell}{\partial y_2(k-1)} \vspace{0.2cm} \\ 
\displaystyle  \frac{\partial F_2^\ell}{\partial y_1(k-1)} &  \displaystyle \frac{\partial F_2^\ell}{\partial y_2(k-1)}
\end{array}
\right]
\end{eqnarray}
evaluated at such fixed points. The  hyperbolic fixed point can be classified as: an
{\it attractor}\, if $|\lambda_i| <1,~\forall i=1,2,\dots,p$; a repellor if $|\lambda_i| >1,~\forall i=1,2,\dots,p$; 
and a {\it saddle}\, if there are eigenvalues inside and outside the unit circle.  A fixed point will be indicated by $\bar{\by}^*$.

As said before, if black-box techniques are used, typically the resulting vector field will not have a fixed
point at the origin. This can be forced upon the model by removing the constant term of the pool of candidate regressors \citep{agu_men/96}.
Unfortunately, although a fixed-point will appear at the origin by imposition, we cannot guarantee that it will
be an attractor, as a matter of fact it will usually be a saddle. The good news is that because the demonstrations 
in $\mathbf{Y}_{(n)}$ all converge to the target, the estimated vector field usually has an attractor
slightly off the origin. This can be understood as a ``perturbed'' version of the ideal solution where the
perturbations come from the imperfections of the trajectories, that are provided not by a vector field but
rather by a human teacher.

A way of circumventing this problem is then to impose a fixed point nearby the origin. In doing so, this new equilibria
will retain its type (a saddle in this example) and the origin becomes an attractor, as desired. Some details are provided in what
follows.  In order to impose that the model  (\ref{a070116}) should have a fixed point $\bar{\by}^*=[\bar{y}_1, \bar{y}_2]^{\T}$,
a constraint of the form ${\rm c}=S\btheta$ with
\begin{equation}
\label{restricao}
{\bf c} = \left[ \begin{array}{c}
\bar{y}_1 \\
\bar{y}_2 
\end{array}
\right],  \qquad
S= \left[ \begin{array} {c}
F^\ell_1[\bar{y}_1,\, \bar{y}_2] \\
F^\ell_2[\bar{y}_1,\, \bar{y}_2]
\end{array} \right]
\end{equation}

\noindent
should be satisfied. This can be achieved using the estimator (\ref{constrained}), as mentioned before.

Let us call ${\cal M}$ an unconstrained model estimated from the data $\mathbf{Y}_{(n)}$
which has a fixed point at $\bar{\by}_1^*$. The parameters can be estimated anew from the same data using (\ref{constrained}) with
(\ref{restricao}) in order to impose a new fixed point at $\bar{\by}^*$. The constrained model is referred to as ${\cal M}_{\rm c}$. 
It has been conjectured that the fixed point $\bar{\by}^*$ of ${\cal M}_{\rm c}$ will be of the same type as that of $\bar{\by}_1^*$ of ${\cal M}$ 
for sufficiently small $|\bar{\by}_1^*- \bar{\by}^*|$ \citep{san_eal/18}.

This procedure has been performed for a whole library of demonstrations (available on the Web) and 
described in  \citep{san_eal/18}. A single example is given below for the sake of illustration.

\begin{figure}
\centering
\begin{tabular}{c}
(a)\\ 
\includegraphics[width=0.45\textwidth]{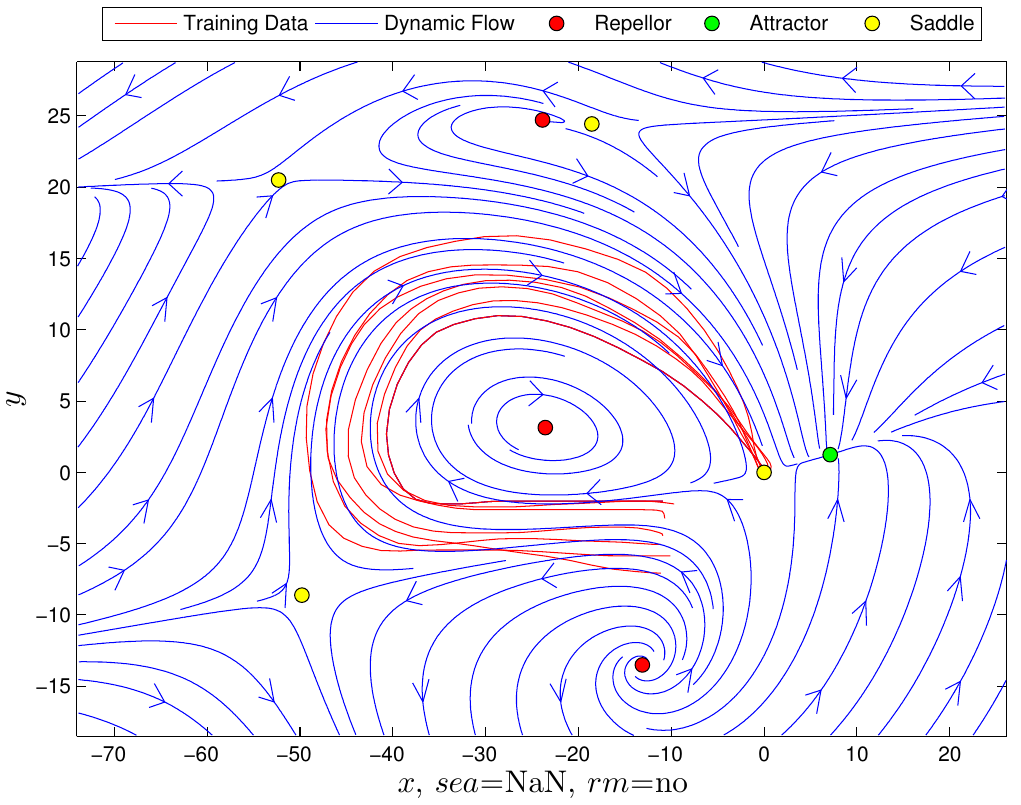} \\ (b) \\
\includegraphics[width=0.45\textwidth]{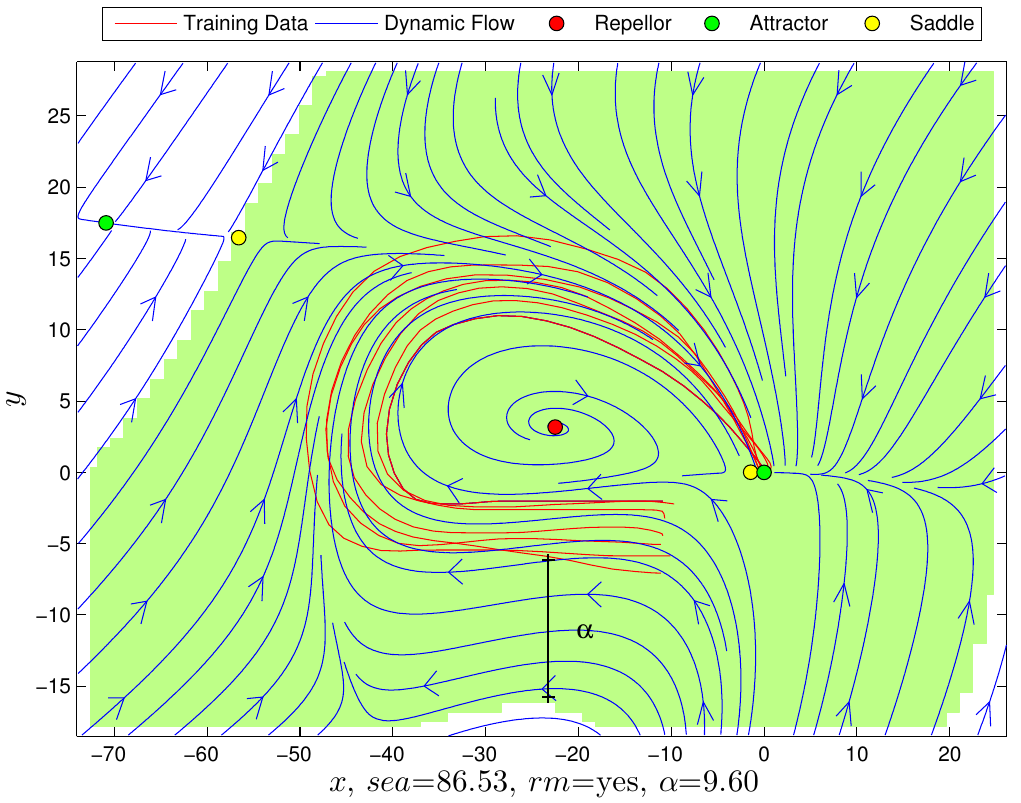}\\ (c)\\
\includegraphics[width=0.45\textwidth]{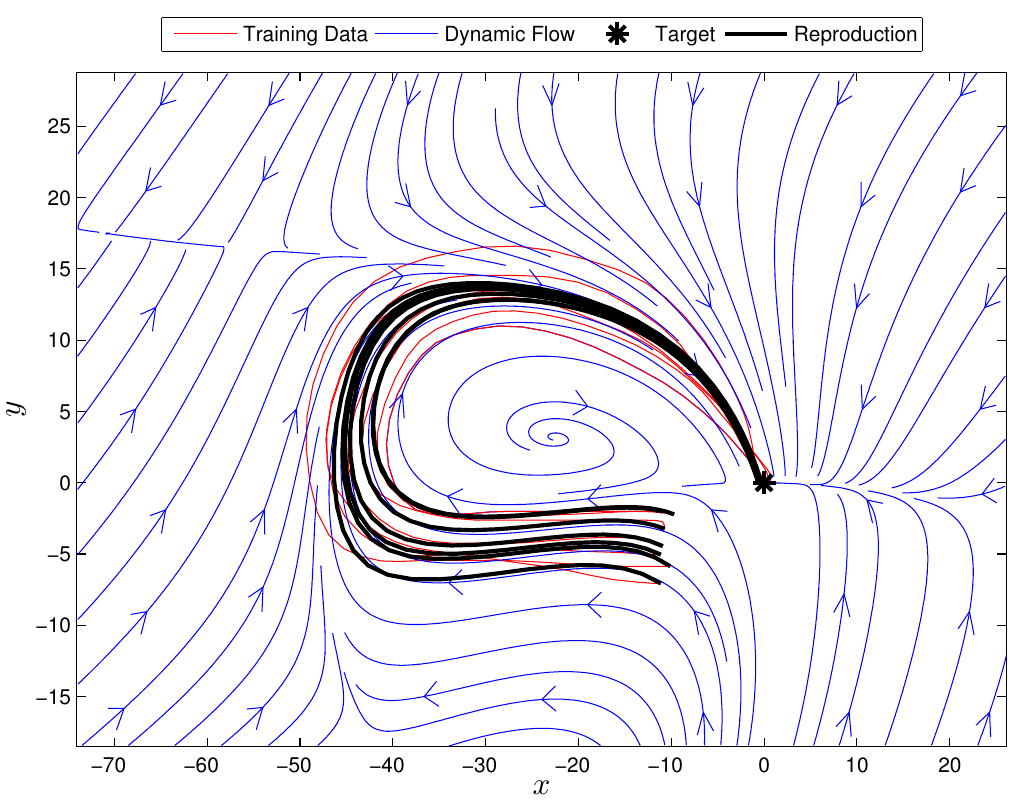} 
\end{tabular}
\caption{\label{fig:caso} \small (a)~The set of teacher-produced demonstrations is shown in red and the blue lines illustrate the
vector field of a black-box model. (b)~Result of the use of grey-box techniques, notice that now the origin
is an attractor with its basing of attraction shown as the green area. (c)~Different initial conditions are used
to produce new model-generated trajectories (in black) using (\ref{a280419}). See text and  \citep{san_eal/18} for details.}
\end{figure}

The starting point is the set of demonstrations shown in red in Figure~\ref{fig:caso}a  and the blue lines illustrate the
vector field of a black-box model. As it can be seen, the fixed point at the origin is a saddle, and there is
an attractor slightly to the right. Using the grey-box techniques described, a fixed-point is imposed very
close (to the left) of the origin, see Figure~\ref{fig:caso}b. When parameters are estimated again, the origin becomes
the attractor. It is as if the saddle, originally located at the origin, would shift to the left where the new fixed-point
is imposed, and the attractor, originally to the right, would take residence at the origin which is guaranteed to be
an equilibria by proper structure selection \citep{agu_men/96}. The  model estimated using grey-box techniques is:
\begin{equation}          
\label{a280419}                                                 
\begin{array}{l}                                                           
\begin{array}{rll}    
 y_1(k) =& +0.983506 \, y_1(k-1) +0.096590 \, y_2(k-1) \\
      &  -0.000078 \, y_1(k-1)^3  +0.005253 \, y_2(k-1)^2 \\          
      &  -0.000538 \, y_2(k-1)^3 -0.016513 \, y_1(k-1) \, \, y_2(k-1) \\
      &  -0.000300 \, y_1(k-1)^2 \, \, y_2(k-1)-0.004126 \, y_1(k-1)^2 
\\                                                                         
\\                                                                         
y_2(k) = & +0.779775 \, y_2(k-1) -0.000042 \, y_1(k-1)^3 \\
     &    -0.015285 \, y_1(k-1) \, \, y_2(k-1)  -0.002493 \, y_2(k-1)^2 \\ 
     &    -0.000216 \, y_1(k-1)^2 \, \, y_2(k-1) -0.004130 \, y_1(k-1) \\
     &    -0.000102 \, y_2(k-1)^3  -0.001130 \, y_1(k-1)^2\\
     &   +0.000001 \, y_1(k-1) \,  \, y_2(k-1)^2\,.  \nonumber     
\end{array}                                                               
\end{array}                                                               
\end{equation}

Figure~\ref{fig:caso}c shows in black some model-produced trajectories, which have all the same type of 
behavior than the original demonstrations produced by the human teacher. It should be noted that a great
variety of trajectories can be produced by taking initial conditions from other regions in $\bbR^2$ within the
basin of attraction.

This procedure has been successfully applied to a small mobile robot and to a Comau Smart Six manipulator.
Details can be found in \citep{san/18tese} and in the videos available at~\url{https://goo.gl/AqMLAH}.

\subsubsection{Benchmarks}

To conclude this section, some pointers will be given for those who are interested in 
benchmarks and datasets. 

The effort and initiative of Maarten Schoukens and Jean-Philippe No\"el is worth pointing
out \citep{sch_noe/17}. The website {\tt https://www.nonlinearbenchmark.org}
should be checked out \citep{sch_noe/16}.

Some of the data sets used by the author in his papers are freely available, as listed below:

\begin{enumerate}
\item chaotic data from Chua's circuit \citep{agu_eal/97}: \\
{\tt https://www.researchgate.net/publication/319493329\_ChuaData1} 

\item data from a buck converter \citep{agu_eal/99i3e}: \\
\begin{footnotesize}
{\tt https://www.researchgate.net/publication/265412913\_Measured\_Data\_of\_Buck\_Converter}
\end{footnotesize}

\item data from a control valve \citep{agu/14}:\\
\begin{footnotesize}
{\tt https://www.researchgate.net/publication/263334571\_DeadZoneValveData\_RG}
\end{footnotesize}

\item dynamical and static data from a heater \citep{agu_eal/02iee,agu_eal/05iee}:\\
{\tt https://www.researchgate.net/publication/260177982\_estat3}\\
{\tt https://www.researchgate.net/publication/260177701\_DIN3}\\
{\tt https://www.researchgate.net/publication/260177800\_Din4}
\end{enumerate}

\section{Further Reading}
\label{fr}

There is a wealth of books that deal with system identification. Most books cover
the theory related to linear systems. Early examples include \citep{eyk/74,hsi/77}. A more 
modern approach with a view to prediction error methods is given by \citep{nor/86,sod_sto/89,lju/99}.
A text devoted to recursive techniques is \citep{lju_sod/83}. Nonlinear techniques are
discussed in more recent texts \citep{nel/00,bil/13book}. 

Interesting discussions about the effects of the input on the parameter estimation 
process can be found in \citep{baz_eal/08,gev_eal/09}. In a different vein, Singh and
co-workers have investigated the design of training and validating data sets from historical
data \citep{sin_eal/19}.

Nepomuceno and Martins have endeavoured to establish a lower bound for free-run simulation
errors of polynomial NARMAX models with chaotic dynamics \citep{nep_mar/16}. Nepomuceno
has also investigated the use of multiobjective optimization techniques for determining the number
of nodes of random neural networks \citep{nep/19}. Hafiz and co-workers have used
evolutionary  techniques to solve multiobjective optimization problems related to structure selection
of NARX polynomial models \citep{haf_eal/19,haf_eal/20}. Related to this, an extensive review on
probably the most popular algorithms based on computational intelligence to solve optimization
problems related to parameter estimation and system identification has been provided in \citep{qua_eal/20}.

Bayma and co-workers proposed
a way of analyzing identified NARX polynomial models in the frequency domain, by means of
nonlinear output frequency response functions \citep{bay_eal/18}. Ferreira and co-workers proposed
the simultaneous use of the normalized RMSE, a coherence-based index and the fourth-order
cross-cumulant index. Such indices are combined to form a scale for model validation \citep{fer_eal/17}.

An interesting approach to the identification of nonlinear systems is the use of piecewise ARX models, where
not only the model parameters but also the model space partitions must be estimated according to
some optimality criterion \citep{bar_eal/18}. In a similar vein, but using neural-fuzzy models, 
Liu and co-workers besides the local model parameters and the location of the model partitions, 
they also estimate the number of partitions \citep{liu_eal/19}. Similar problems have been
investigated by Barreto and co-workers using self organizing maps and $K$ nearest neighbors algorithms
to partition the data \citep{jun_eal/15,bar_sou/16}. In particular, in \citep{jun_eal/15} the 
concept of regional models is used as a model class which falls between global and local models. A similar
approach which is not limited to local linear models is the Switched NARX or SNARX models. The
challenge of defining the partitions (switching pattern) and structure selection of the local models has been 
addressed by Federico Bianchi and co-workers \citep{bia_eal/20,bia_eal/21}.

Early works on grey-box system identification in the sense treated in this paper include
\citep{esk_eal/93,tul/93,boh/94,joh/96}.

Ribeiro and co-workers have provided insights into the trade-off between the smoothness of the
cost function and i)~the memory retention capabilities during training of neural networks \citep{rib_eal/19},
and ii)~the window over which the power of free-run simulation errors is minimized \citep{rib_eal/20}.

The use of prior information in the context of neural networks is somewhat more involved. 
Some early results have been reported in \citep{tho_kra/94,ama/01,agu_eal/04,li_pen/06,fre/13}.
Grey-box identification techniques for radial basis function (RBF) networks has been
addressed in \citep{agu_eal/04,agu_eal/07,che_eal/11}.

\cite{bar_eal/11} used the uncertainty related to sensor data to choose models from the
Pareto set in biobjective optimization. A number of model structures were compared, including
neural networks using data from hydraulic pumping system. 
Also, two different cost functions related to the dynamical data set were
used one at a time: the conventional $J(Z,\,Z_{{\cal M}_1})$ which results in a convex optimization problem
and $J(Z,\,Z_{{\cal M}_{\rm s}})$, which results in a nonconvex optimization problem, that was
solved using genetic algorithms. The objective related to static data was always $J(Z_{\rm ss},\,Z_{{\cal M}_{\rm ss}})$.

The use of biobjective estimation has been used to handle model uncertainty in the parameters
\citep{tei_agu/10} and in the model structure \citep{bar_eal/15}. As a matter of fact, the
problem of characterizing and dealing with structural uncertainty remains an open problem
that has attracted attention \citep{bal_eal/13,gu_wei/18}. A biobjective cost function was used
in \citep{mav_eal/20} in the context of neural network training. The second objective did not
convey any auxiliary information as such, but was used as a regularization term.

In \citep{wu_eal/20} two different data sets are used to train two submodels. The combination of
the outputs of both submodels form the final output. The parameters are estimated by minimizing 
a multiobjetive cost function that penalizes the error of the final model, the difference between
the submodel outputs and the number of parameters.

Using the evidences of symmetry in the dynamics of the solar dynamo \citep{let_eal/06}, the
time series of the sunspots was mapped unto a symmetric space where it was possible to
build models -- that used symmetry as auxiliary information -- to forecast the time series \citep{agu_eal/08solar}.

Wei and Billings have discussed the modeling of COVID-19 dynamics and aspects of interpretability
of NARMAX models \citep{wei_bil/21}. In this interesting application the use of large lags for both
input and output, due to the delays associated with the pandemic dynamics, seem to be a keep aspect
of the modeling. An interpretation of the use of large lags related to cyclical data has been put
forward in  \citep{agu_eal/08jepe}.

\section*{Acknowledgements}

Financial support by CNPq and FAPEMIG (Brazil) is gratefully acknowledged. The careful
reading of this manuscript by Petrus Abreu is acknowledged with thanks.

\bibliographystyle{apalike}

\begin{small}


\end{small}

\end{document}